\documentclass[conference]{IEEEtran}
\IEEEoverridecommandlockouts
\usepackage{cite}
\usepackage[inline]{enumitem}
\usepackage{amsmath,amssymb,amsfonts}
\usepackage{subfig}
\usepackage{graphicx}
\usepackage{textcomp}
\usepackage{xcolor}
\usepackage{soul}
\usepackage{caption}
\usepackage{epstopdf}
\usepackage{array}
\usepackage{nicefrac}
\usepackage{siunitx}
\usepackage{adjustbox}
\usepackage[utf8]{inputenc}
\DeclareUnicodeCharacter{FB01}{fi}
\usepackage[colorinlistoftodos]{todonotes}
\makeatletter

\newcommand{\inlineitem}[1][]{%
\ifnum\enit@type=\tw@
    {\descriptionlabel{#1}}
  \hspace{\labelsep}%
\else
  \ifnum\enit@type=\z@
       \refstepcounter{\@listctr}\fi
    \quad\@itemlabel\hspace{\labelsep}%
\fi}
\makeatother
\def\BibTeX{{\rm B\kern-.05em{\sc i\kern-.025em b}\kern-.08em
    T\kern-.1667em\lower.7ex\hbox{E}\kern-.125emX}}

\newcommand{\comment}[1]{ }

\usepackage{titlesec}
\titlespacing*{\section}
{0pt}{1.2ex}{0ex} 
\titlespacing*{\subsection}
{0pt}{1ex}{0ex}  
\titlespacing*{\subsubsection}
{0pt}{0ex}{0ex} 

\begin{document}











\title{Leveraging Hardware-Impaired Out-of-Band Information Through Deep Neural Networks for Robust Wireless Device Classification}

\author{\IEEEauthorblockN{Abdurrahman Elmaghbub and Bechir Hamdaoui}
\IEEEauthorblockA{\textit{School of Electrical Engineering and Computer Science} \\
\textit{Oregon State University}\\
\{elmaghba, hamdaoui\}@oregonstate.edu}

}

\maketitle
\begin{abstract}
Wireless device classification techniques play a key role in promoting emerging wireless applications such as allowing spectrum regulatory agencies to enforce their access policies and enabling network administrators to control access and prevent impersonation attacks to their wireless networks. Leveraging spectrum distortions of transmitted RF signals, caused by transceiver hardware impairments created during manufacture and assembly stages, to provide device classification has been the focus of many recent works. These prior works essentially apply deep learning to extract features of the devices from their hardware-impaired signals, and rely on feature variations across the devices to distinguish devices from one another.
As technology advances, the manufacturing impairment variations across devices are becoming extremely in significant, making these prior classification approaches inaccurate.
This paper proposes a novel, deep learning based technique that provides scalable and highly accurate classification of wireless devices, even when the devices exhibit insignificant variation across their hardware impairments and have the same hardware, protocol, and software configurations.
The novelty of the proposed technique lies in leveraging both the {\em in-band} and {\em out-of-band} signal distortion information by oversampling the captured signals at the receiver and feeding IQ samples collected from the RF signals to a deep neural network for classification. Using a convolutional neural network (CNN) model, we show that our proposed technique, when applied to high-end, high-performance devices with minimally distorted hardware, doubles the device classification accuracy when compared to existing approaches. 
%

\end{abstract}

\section{Introduction}

Wireless device classification techniques based on hardware impairments have attracted considerable amount of attention in recent years. These techniques emerge as key enablers to some important spectrum/network access awareness applications. They, for example, allow spectrum regulatory agencies to enforce their spectrum access and sharing policies, and enable network administrators to prevent impersonation attacks and unauthorised access to their wireless networks. 
Hardware-impairment based classification techniques essentially leverage distortions in transmitted RF signals that are caused by manufacturing impairments in the transceiver hardware to distinguish among devices. They do so by employing deep learning methods that automate feature extraction and device classification from IQ samples collected from the received RF signals. These device-specific hardware impairments embedded during manufacturing and assembly processes provide unique device signatures that cannot be easily cloned, spoofed or modified, thereby allowing to detect and prevent impersonation attaches in addition to increasing device classification accuracy. 

Due to its automated decision-making nature, deep learning has recently been used to develop high-performing device classification methods. However, the training/testing accuracy of these deep learning based methods decreases with the decrease of the impairment variability among the wireless devices. Therefore, using these methods, it can be very difficult to distinguish and separate among devices with very similar distortion values. In addition, these device signatures become more vulnerable to environment distortions such as wireless channel fading and system noise. For instance, high-end, bit-similar software-defined radios (SDRs), such as USRP X310 radios, are composed of low-variability components that render them not easy to identify using existing deep learning based methods. Oracle~\cite{sankhe2019oracle}, on the other hand, intentionally introduces impairments in the signal to increase the differentiability among devices while maintaining a tolerable bit error rate (BER) for each device. DeepRadioID~\cite{2019deepradioid} also leverages a carefully-optimized digital finite response filter (FIR) at the transmitter's side to slightly modify the baseband signal to compensate for current channel condition. These methods showed considerable improvement and resiliency against high similarity among transmitters and high channel condition variability. However, they suffer from scalability issues, since the set of artificial impairment values to be added before exceeding the tolerable BER level is limited. Additionally, it is not practical to integrate an FIR filter into each transmitter's circuit that desires to interact with the network. 

In this paper, we propose a novel, deep learning-based device classification technique that uses IQ samples collected from the RF signals to efficiently identify and classify high-performing transmitters that have the {\em same, minimally-distorted} hardware components. 
The proposed technique: (1) is scalable in that it can distinguish among large numbers of minimally-distorted devices with same hardware, regardless of their protocol/software configurations, 
(2) is robust against signature cloning and modification,
(3) requires no changes at the transmitters, and
(4) incurs minimal extra processing at the receiver side that can be performed with existing hardware. 
The novelty of the proposed technique lies in considering both the {\em in-band} and {\em out-of-band} spectrum emissions of the received signal to capture hardware signatures and features that can be used to uniquely and efficiently discriminate among devices, even when devices have same hardware with significantly reduced distortions. We describe the impact of hardware impairments caused by various analog RF components on {out-of-band spectrum distortions}, and illustrate how such distortions could serve as efficient ways for providing unique signatures of wireless devices. We also show, using simulations, that the proposed technique significantly outperforms existing approaches in terms of classification accuracy, when considering high-performing devices with minimally-distorted hardware components.

The rest of the paper is organized as follows. Section \ref{HW impairments} describes  models and impact of hardware transmitter impairments. Section \ref{ProposedFramework} presents the main idea of the proposed technique, as well as describes the out-of-band distortions arising from the transmitter hardware components that are key contributors to these out-of-band distortions. The results and performance analysis of the proposed technique are described in Section \ref{Performance_eval}. Finally, the conclusion is drawn in Section \ref{Conclusion}.

\section{Transmitter Hardware Impairments}
\label{HW impairments}
\begin{figure*}
    \centering
    \includegraphics[width=1.25\columnwidth]{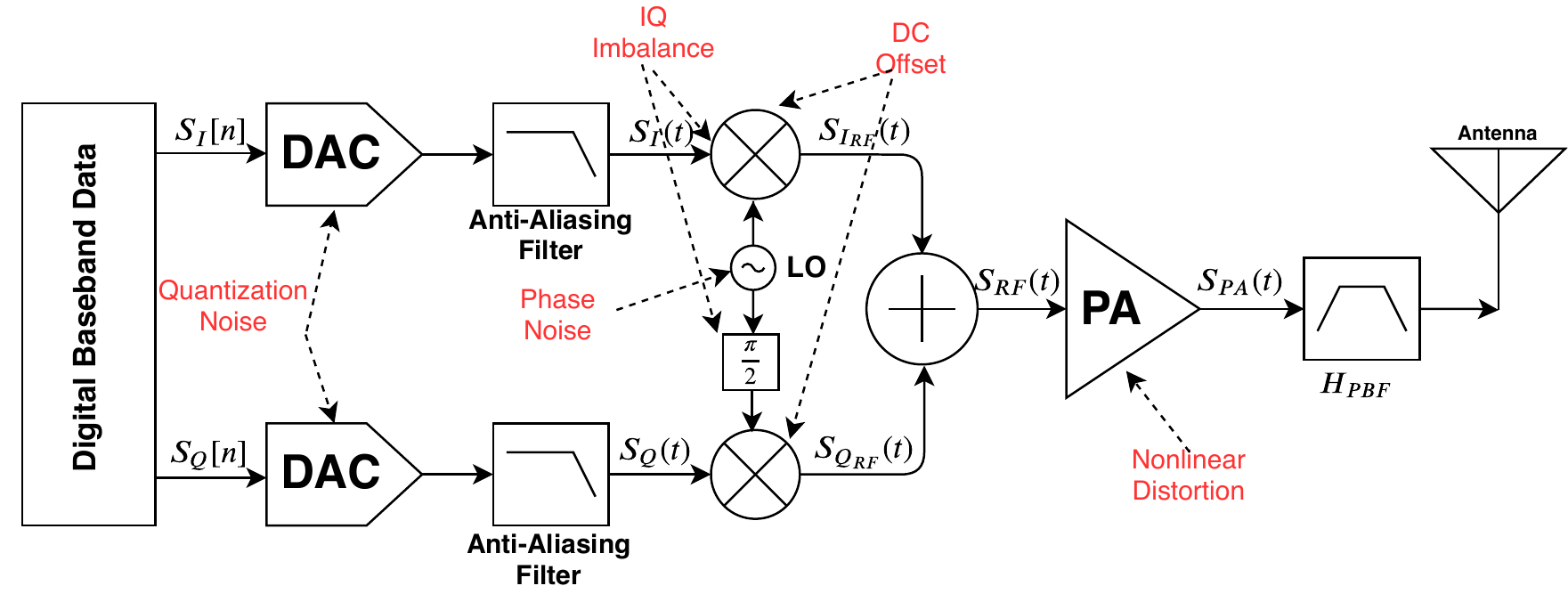}
    \caption{Typical transceiver with various RF impairments}
    \label{fig:tx}
\end{figure*}
RF transmitters acquire benign hardware impairments during manufacturing and assembly stages. These device-specific impairments cause the transmitted RF signals to deviate from their ideal values, thereby establishing unique signatures for their corresponding devices. Despite the many great efforts aimed at designing hardware techniques that can eliminate/limit these hardware impairments so that they fall within tolerable ranges, these impairments cannot be eliminated completely. Therefore, since our focus in this paper is on exploiting such impairments to provide and enable efficient device classification, we begin in this section by taking a closer look at the sources, modeling, and impact of the most significant transmitter-specific impairments. Fig.~\ref{fig:tx}, showing these impairments, will be used throughout for illustration.

\subsection{Quantization Noise and Clock Source Modulation}
Starting our analysis from the interface between the digital processing unit and the analog front-end components in modern transmitters, Digital to Analog Converters (DACs) convert digital baseband sequences (e.g., $S_I[n]$ for the in-phase (I) path component) to their equivalent time-continuous analog signals (e.g., $S_I(t)$). DACs use different variations of zero-order-hold circuits to generate staircase continuous waveforms as an approximation of the smooth waveforms. The high-frequency components represented by the sharp edges of the staircase pattern are removed by the Anti-Aliasing filter. The limited number of DAC resolution bits, the finite clipping levels, and the nonlinearity nature of real DACs~\cite{d2010modeling,smaini2012rf} altogether result in the degradation of the SNR values and the transmitter performance in general. The main three sources of a DAC's distortions are, horizontal quantization (HQ), vertical quantization (VQ), and clock source modulation (CM), and whose aggregated impact on the DAC functionality can be modelled as additive terms superimposed on the ideal analog output. 
For instance, considering the input $S_I[n]$, the DAC output $y(S_I[n])$ in the I path can be modelled as~\cite{d2010modeling}:
\begin{equation}
    y(S_I[n]) = S_I(t) + y_{S_I}^{\rm VQ}(t) + y_{S_I}^{\rm HQ}(t) + y_{S_I}^{\rm CM}(t)
    \label{eq8}
\end{equation}
where $y_{S_I}^{\rm HQ}(t)$, $y_{S_I}^{\rm VQ}(t)$, and $y_{S_I}^{\rm CM}(t)$ represent HQ, VQ, and CM distortions, respectively. HQ distortion represents the built-in discrete nature of the DAC output since it produces staircase patterns by holding the sample value during the sampling period. Anti-Aliasing filters are used to eliminate the superimposed frequency components due to this discrete nature; however, interfering spurious terms still appear closer to the bandwidth of the output signal when the generating frequency is not sufficiently greater than the Nyquist rate. This effect can be modelled as~\cite{d2010modeling}
\begin{equation}\nonumber
    y_{S_I}^{\rm HQ}(t) = \sum_{n = -\infty}^{\infty}S_I[n]g\left(\frac{t - nT_{g}}{ T_{g}}\right) - S_I(t)
\end{equation}
where $g(\theta)$ is a unitary pulse with $0\leq\theta<1$ and $T_g$ is the generation time period.
Note that in this section and throughout the paper, we use the in-phase (I) path component as an example to illustrate and explain the presented concepts. Similar analysis and illustration could of course be done for the case of the quadrature (Q) path component.

The time-domain instability of the clock source is what leads to periodic variation in the generating period, resulting in the CM impairment, which in turn generates unwanted spurious components in the signal spectrum and can be modeled as~\cite{d2010modeling}
\begin{equation}\nonumber
    y_{S_I}^{\rm CM}(t) = \sum_{n = -\infty}^{\infty}S_I[n]h_{n}(t - nT_{g})
\end{equation}
where the function $h_n(t)$ is defined as~\cite{d2010modeling}
\begin{equation*}
   {h_{n}(t) \!\!=\!\! -{\rm sign}(\Delta_{n})g\!\! \left(\!\!\frac{t - nT_{g}}{ \Delta_{n}}\!\!\right) \!\! +\!\! \ {\rm sign}(\Delta_{n + 1})g\!\! \left(\!\!\frac{t - \left(n + 1\right)T_{g}} {\Delta_{n + 1}}\!\!\right)}
\end{equation*}
%
where $sign(\theta)$ is the sign function, and $\Delta_n$ is the deviation of the clock from its ideal value. 

The finite resolution of the DAC requires rounding the samples values to the nearest voltage level, referred to above as vertical quantization or VQ, giving rise to a quantization distortion proportional to the DAC resolution. Similar to the previous DAC impairments, VQ distortion increases the spurious content in the spectrum as well. It can be modelled as~\cite{d2010modeling}
\begin{equation*}
y_{S_I}^{VQ}\!(t) \!\!=\!\!\!\!\!\!\! \sum_{n = -\infty}^{\infty} \!\!\!\!\! \big(\!\hat{S_I}[n] + T[\hat{S_I}[n]] - S_I[n]\big) \! \big(g(\frac{t - nT_{g}}{T_{g}})   + h_{n}(t - nT_{g})\big)
\end{equation*}
where $T[\hat{S_I}[n]]$ is the Integral Nonlinearity (INL) term, which is a measure of the deviation of the output values from the ideal, and $\hat{S_I}[n]$ is the approximated signal values. In ideal DAC, $T[\hat{S_I}[n]] = 0$, making the VQ term go to zero. 

Each of these three DAC distortions (VQ, HQ, and CM) is hardware dependent, and hence, can be exploited as a feature/signature to distinguish one transmitter from another.

\subsection{IQ Imbalance} 
Zero Intermediate Frequency (IF) or direct transmitters, such as the one shown in Fig.~\ref{fig:tx}, leverage the quadrature mixer configuration to implement the upconversion of the baseband signal without the need for using any filtering methods. 
It does so by separately (in parallel) upconverting, at the carrier frequency $w_c$, the two in-phase (I) baseband modulated component, $S_I(t)=A(t)\cos(\phi(t))$, and quadrature (Q) baseband modulated component, $S_Q(t)=A(t)\sin(\phi(t))$, using two independent mixers fed by a local oscillator (LO) tone shifted by 90\si{\degree} from one another. 
%
%
Assuming perfectly matched I and Q paths, the two outputs are summed up, yielding the bandpass modulated signal (see Fig.~\ref{fig:tx})
\begin{eqnarray} 
  S_{RF}(t) \!\!\!\!&=&\!\!\!\! A(t)\cos(\phi(t))\cos(w_ct) - A(t)\sin(\phi(t))\sin(w_ct) \nonumber 
\end{eqnarray}

Any amplitude mismatch $\Delta\alpha$ or phase deviation $\Delta\theta$ between the I and Q path components that can be caused by DAC and/or mixer hardware impairements lead to imperfect image cancellation and result in residual energy at the mirror frequency $-w_c$, causing interference and SNR degradation. This amplitude mismatch and phase deviation, aka IQ imbalance, can be quantified by measuring the power ratio between the image and the desired signal, which depends on $\Delta\alpha$ and $\Delta\theta$. 

When using {\em real} mixers with amplitude and phase imbalances of $\Delta\alpha$ and $\Delta\theta$, 
the upconverted (distorted) signal becomes:
\begin{equation}
{S}_{RF}(t) = (1-\Delta \alpha)S_{I}(t)\cos(w_ct) - S_{Q}(t)\sin(w_ct  + \Delta\theta) \nonumber
\end{equation}
Now when masking all other hardware impairments (i.e., assuming all other hardware components are ideal except DACs and mixers), the distorted complex baseband signal $\tilde{R}(t) = {S}_{RF}(t)e^{-jw_ct}$ down-converted at the receiver is 
\begin{equation*}
\tilde{R}(t)=[(1-\Delta \alpha)S_{I}(t)\cos(w_ct) - jS_{Q}(t)\sin(w_ct + \Delta\theta)]\times
\end{equation*}
\begin{equation}\label{eq:drx}
    [cos(w_ct) - j\sin(w_ct)]
\end{equation}
After some math manipulations and clearing the terms appearing at twice the carrier frequency (which will be filtered out via low-pass filtering at the receiver), Eq.~\eqref{eq:drx} yields
\begin{equation}
\tilde{R}(t) \! =\! \bigg(\!\frac{1-\Delta \alpha}{2}\!\bigg)S_I(t) \!+\! j \bigg(\!\frac{\sin(\Delta\theta) - j \cos(\Delta\theta)}{2}\!\bigg) S_Q(t) \nonumber
\end{equation}
Therefore, IQ imbalances manifest in in-band and out-of-band signal distortions that can be exploited to increase device signature separability and classification accuracy.
 
\subsection{DC Offset}
Ideal mixers output the product of the two signals coming from the input and the LO ports, which consists of only two terms, one appearing at the summation of the multiplied frequencies and one at their subtraction. However, due to hardware impairments, real mixers  also produce some other unwanted emissions at different frequencies. Of a particular importance is a spike that appears at the center of the desired signal spectrum, known as DC offset, which cannot be easily filtered out because of its location in the middle of the message spectrum. 
DC offset impairments distort signal constellations and increase the error vector magnitude.  

There are two main sources of DC offsets: carrier leakage and second-order nonlinearity. Carrier leakage results from the LO leakage coming from the poor isolation between the three mixer ports limited by the different coupling effects. Thus, a strong LO signal can leak through unintended paths toward the mixer output port and appear at the middle of the desired signal spectrum, generating a static DC value at the receiver~\cite{svitek2005dc}. For example, when mixing the in-phase component $S_I(t)$ while considering this LO leakage, the mixer output becomes~\cite{svitek2005dc} 
$$S_{I_{RF}} = S_I(t)\cos(w_ct) + v_{lo}\cos(w_ct),$$ 
where $v_{lo}\cos(w_ct)$ is the unmodulated carrier term that leaks through the mixer output port and appears at the middle of the spectrum. $v_{lo}$ is a hardware-specific feature that varies from a mixer to another.

The second source of DC offset is second-order nonlinearity. When a single tone signal passes through a second-order nonlinearity system, the output signal exhibits frequency components at the integer multiple of the input frequency. To illustrate, consider feeding the in-phase baseband component to the mixer while considering only the nonlinearity up to the second-order and ignoring the LO leakage effect. The output of the mixer in this case becomes $S_{I_{RF}}(t)= \alpha_1 S_I(t)\cos(w_ct) + \alpha_2 S^2_I(t)\cos^2(w_ct)$, where $\alpha_1$ and $\alpha_2$ are the parameters that model the mixer's first- and second-order nonlinearity terms.  
When replacing $S_I(t)$ by its expression $A(t)\cos(\phi (t))$, the second-order nonlinearity term---the one responsible for the DC component---can be written as
\comment{\begin{equation}\nonumber
    y(S_I(t)) = \frac{\alpha_2 A(t)^2}{2} + \alpha_1A(t)\cos(\phi (t)) + \frac{\alpha_2A^2(t)}{2}\cos(2\phi(t))
\end{equation}}
\begin{equation*}\nonumber
    \alpha_2 S^2_I(t)\cos^2(w_ct) = \frac{\alpha_2 A^2(t)}{4} + \frac{\alpha_2 A^2(t)}{8} \big[2\cos(2w_ct)+
\end{equation*}
\begin{equation}
    2\cos(2\phi(t))+\cos(2(\phi(t)-w_ct))+\cos(2(\phi(t) + w_ct))\big]
    \label{eq:3}
\end{equation}
Note that the first term in Eq.~\eqref{eq:3} represents the DC component, and it is affected by the nonlinearity distortion captured by the parameter $\alpha_2$. The characteristics of the DC component are determined by both the silicon-level circuitry of the LO and the second-order nonlinearity of the device. Therefore, DC offsets also contribute to the establishment of unique signatures and hardware features that can be leveraged for uniquely identifying transmitters among one another.

\subsection{Phase Noise}
\begin{figure}
     \subfloat[Ideal Local Oscillator.\label{subfig-1:phase1}]{%
       \includegraphics[width=0.235\textwidth, height = 0.18\textwidth]{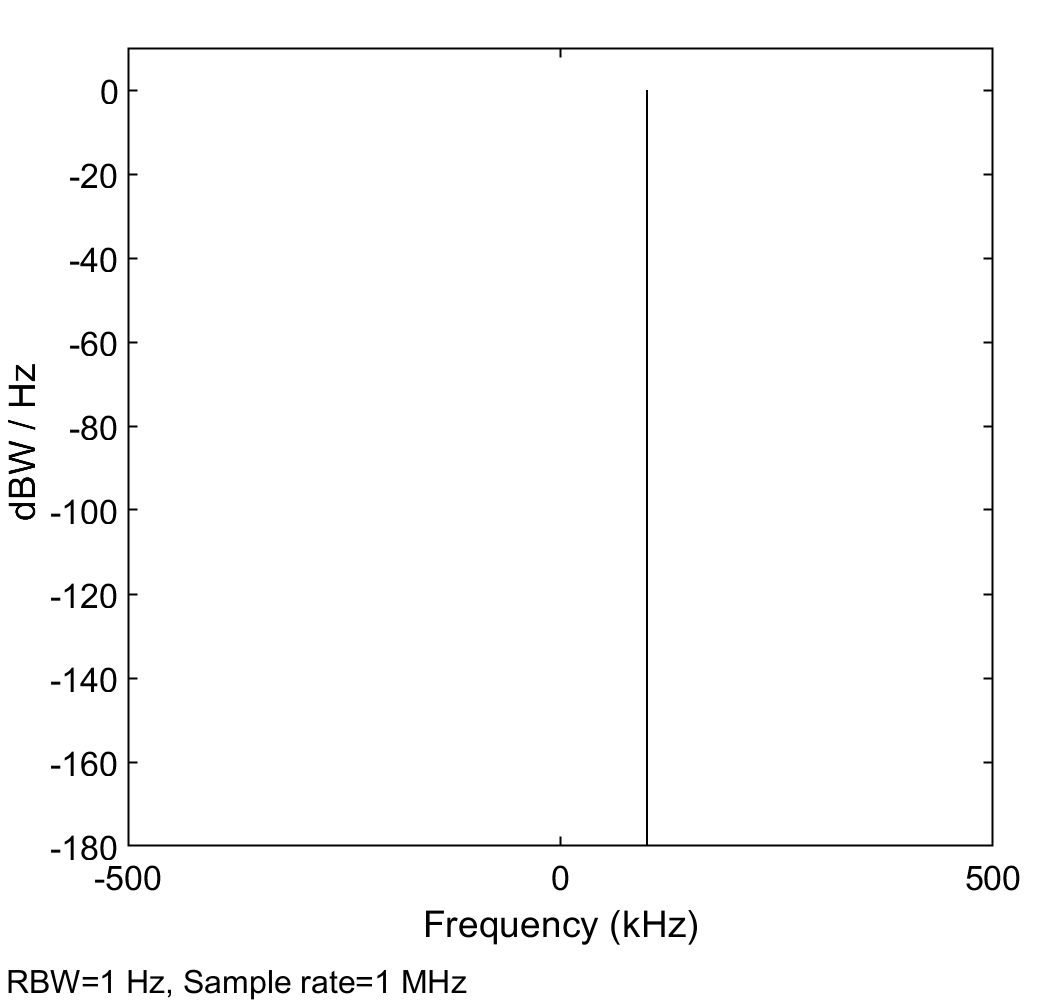}
     }
    \hspace{0.00001cm}
     \subfloat[Real Local Oscillator.\label{subfig-2:phase2}]{%
       \includegraphics[width=0.235\textwidth, height = 0.18\textwidth]{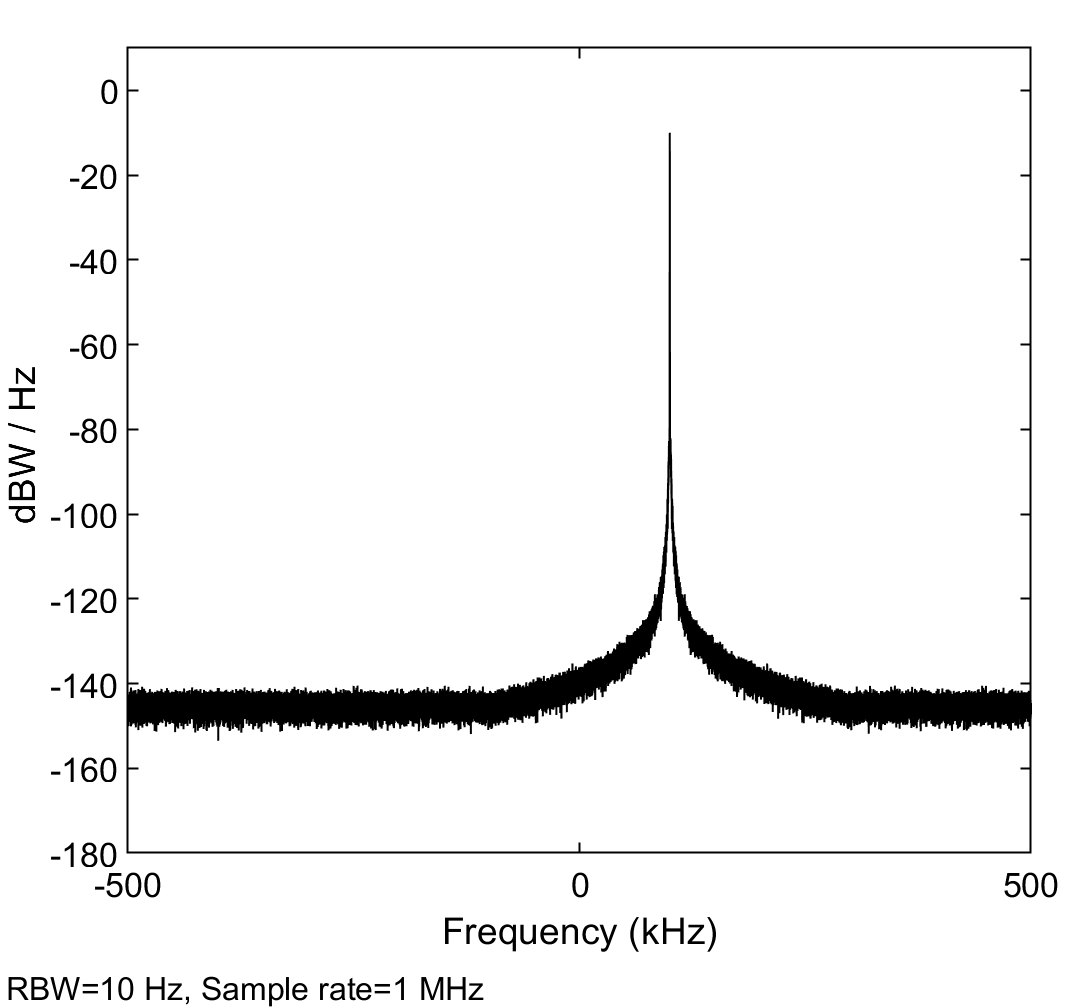}
     }
     \caption{Phase Noise Effect}
     \label{fig:ph}
\end{figure}

Local oscillators (LOs) are fundamental blocks in RF transmitter architectures. They are mainly responsible for producing periodic oscillating signals that can be used by the mixer to upconvert the baseband signal at the carrier frequency. The output of an ideal LO can be represented as a pure sinusoidal waveform $\cos(w_ct)$ that would help to translate signals to the RF domain while preserving the original spectrum shape. Fig.~\ref{subfig-1:phase1} shows the upconversion of a baseband tone to $100$KHz using an ideal LO signal. Similar to the clock source issue in the DAC, the time domain instability of the generated signals by real LOs causes random phase fluctuations, known as Phase Noise, that expand the signal spectrum by introducing unwanted spectrum in both sides of the carrier frequency. This can be seen in Fig. \ref{subfig-2:phase2} which shows the same previous frequency translation (Fig.~\ref{subfig-1:phase1}), but using a real LO signal, which can be represented as $\cos(w_ct + \theta(t))$ with $\theta(t)$ being the phase noise term .

The phase noise manifests in different noises within the LO circuit, such as thermal noise and flicker noise. It can be quantified by measuring the power of the 1-Hz bandwidth at a frequency offset with respect to the carrier frequency. It results in a random rotation in the receiver signal constellation, thereby increasing the symbol detection error~\cite{khanzadi2014calculation} as well as the out-of-band noise level. To show this impact, consider mixing the in-phase baseband signal, $S_I(t)$, with an LO signal, $\cos(w_ct + \theta(t))$. After upconversion, the mixer output can be expressed as
\begin{equation}
    S_{I_{RF}}(t) = S_I(t)\cos(w_ct+ \theta(t)) = S_I(t)\Re(e^{jw_ct}e^{j\theta(t)})
    \label{eq:7}
\end{equation}
where here $e^{j\theta(t)}$ is the phase noise term, and $\Re(x)$ refers to the real part of complex $x$. 
Given $\theta(t)$ is small and using the approximation $e^{j\theta(t)} \approx1+j\theta(t)$, Eq.~\eqref{eq:7} can be rewritten as
\begin{equation}
    S_{I_{RF}}(t) \approx S_I(t) \cos(w_ct) - S_I(t) \theta(t)\sin(w_ct)
    \label{eq:pn}
\end{equation}
We can see from Eq.~\eqref{eq:pn} that the transmitted signal is composed of an undistorted component and a LO-dependent, phase-noise distorted component of the upconverted signal. 
This LO-dependent component implies that phase noise can also be considered as one of the hardware impairments that can contribute to transmitters' signatures, and hence can be leveraged to increase device distinguishability.

\subsection{Power Amplifier (PA) Nonlinearity Distortion}
The majority of circuit nonlinearity is attributed to PAs, which are the last elements in the transmitter chain. They provide the modulated RF signals with the required radiation power to reach their destination. When a PA operates in the linear region, its I/O characteristics is deterministic and an acceptable performance is ensured. However, operating in that region leads to more power consumption due to the associated low-power efficiency characteristic. Since PAs are major power-hungry blocks, most of the transmitters drive their PAs to work near the saturation region to be more power efficient. Unfortunately, power efficiency and linearity of the PA conflict one another. Hence, the signal would be severely suffered from the nonlinearity of the PA when it works in the saturation region. The nonlinearity distortion results in amplitude compression, as well as in high adjacent channel power as a result of the bandwidth expansion, known as spectral regrowth. Although many linearization methods have been proposed to minimize the distortion and attenuate the spectral regrowth, PAs still exhibit some nonlinearity. 

PA nonlinearity distortion is typically captured through the instantaneous amplitude and phase responses to changes in the amplitude of the input signal, respectively known as Amplitude-to-Amplitude (AM-AM) and Amplitude-to-Phase (AM-PM) distortion curves. Using the complex power series~\cite{blachman1964band} to model the bandpass nonlinearity of the PA, the PA output ${S}_{PA}(t)$ that models the instantaneous AM-AM and AM-PM distortions can be expressed as:
\begin{equation}
    {S}_{PA}(t) = \sum_{n=0}^{\frac{N-1}{2}} \frac{\Tilde{\alpha}_{2n+1}} {2^{2n}}\bigg\{{2n\!+\!1\choose n\!+\!1}\big[(A(t)^{2n}\Tilde{S}(t)\big]\bigg\} e^{j\omega_{c}t}
    \label{eq:5}
\end{equation}
%
where $\Tilde{\alpha_i}$s are the complex coefficients of the model, $N$ is the maximum order of nonlinearity, \!\!\!\!
and $\Tilde{S}(t)=S_I(t)+jS_Q(t)$ is again the complex baseband envelope of the signal.
As we can infer from Eq.~\eqref{eq:5}, only the odd
terms can be determined from single-tone complex compression characteristics, but fortunately, the odd-order terms are
the most important as they produce intermodulation distortion
in-band and adjacent to the desired signal \cite{gard1999characterization}.
Each nonlinear RF components enjoys a variation of I/O characteristics, leading to a unique distortion that is captured by a unique set of coefficients $\Tilde{\alpha_i}$, and can therefore help in composing the device's unique signature~\cite{wang2016wireless}.

\section{Leveraging Out-of-Band Distortions for Robust Device Classification}
\label{ProposedFramework}
\subsection{The Proposed Technique}
Out-of-Band (OOB) emissions are the emissions in the frequencies immediately outside the message bandwidth that predominate the OOB domain. OOB domain is defined as the frequency range separated from
the assigned frequency of the emission by less than 250\% of the message bandwidth\cite{tanaka2008unwanted}. These emissions are mainly caused by the modulation and the nonlinear components of an RF transceiver front-end, and result in in-band distortions as well as in an interference into adjacent channels. As a result, spectrum regulatory agencies, such as FCC and ITU, specify and regulate the permissible levels of the OOB emissions of different emission classes using OOB spectral masks. 

In variable-envelope modulation schemes (like 16QAM), the spectrum of a modulated signal expands into adjacent channels when it passes through nonlinear components, resulting in an increase in the OOB emissions due to the spectral regrowth~\cite{zhou2002predicting}. The characteristics of a spectral regrowth are directly related to the unique coefficients of the corresponding nonlinear components in the RF transceiver chain. The DAC impairments, also, can generate OOB emissions due to the quantization and clipping noise~\cite{smaini2012rf} as explained in the previous section. The other major RF front-end component that contributes to the OOB emissions is the LO. Due to the phase noise that is impaired with the LOs, these OOB emissions cause both an in-band and out-of-band noise scaled by the signal power. Interestingly enough, the out-of-band spectrum of a Phase-Locked-Loop (PLL), which is a widely used block for frequency synthesis in application-speciﬁc IC designs, is a function of the Voltage-Controlled Oscillator (VCO) parameters~\cite{smaini2012rf}. Despite the endless effort to reduce the OOB emissions via various techniques~\cite{morgan2006generalized,ding2004digital}, there will always be some inevitable amount of the OOB emissions that can be tolerated by standards, but also can be exploited for providing unique device signatures. The novelty of our proposed device classification technique lies in exploiting such OOB emissions to provide accurate and robust classification.

Based on the aforementioned discussion about the relationship between the out-of-band emissions and the hardware impairments of RF front-end components, and the observations we made from our simulation studies, we would be missing valuable indicative information if we process and leverage only the (in-band) message bandwidth for providing device signatures. 
Therefore, we propose in this paper to consider both the in-band and out-of-band spectra by oversampling the captured signals at the receiver with an appropriate factor. Without any further processing, the raw IQ values obtained from the oversampled captured signals are then fed into a deep neural network to provide device identification and classification. In the proposed framework, we use a Convolutional Neural Network (CNN), which has been designed and tuned to recognize devices signatures and identify wireless devices. 
It is also worth mentioning that technology advancements of transceiver designs nowadays (e.g., software defined and cognitive radios) can easily allow for sampling the captured signals in the out-of-band region, and therefore, the proposed technique can be implemented without requiring newly/sophisticated receiver designs.



\subsection{Hardware-Impaired OOB Emissions: Model and Impact}
In this section, we provide more depth and insights on out-of-band (OOB) spectrum distortions that arise from the transmitter hardware components that contribute significantly to these OOB distortions: power amplifier (PA), local oscillator (LO), mixers, and digital-to-analog converter (DAC).

\subsubsection{Power Amplifier (PA)}
\begin{figure}
     \subfloat[Linear PA\label{subfig-1:FSK1}]{%
       \includegraphics[width=0.22\textwidth, height = 0.15\textwidth]{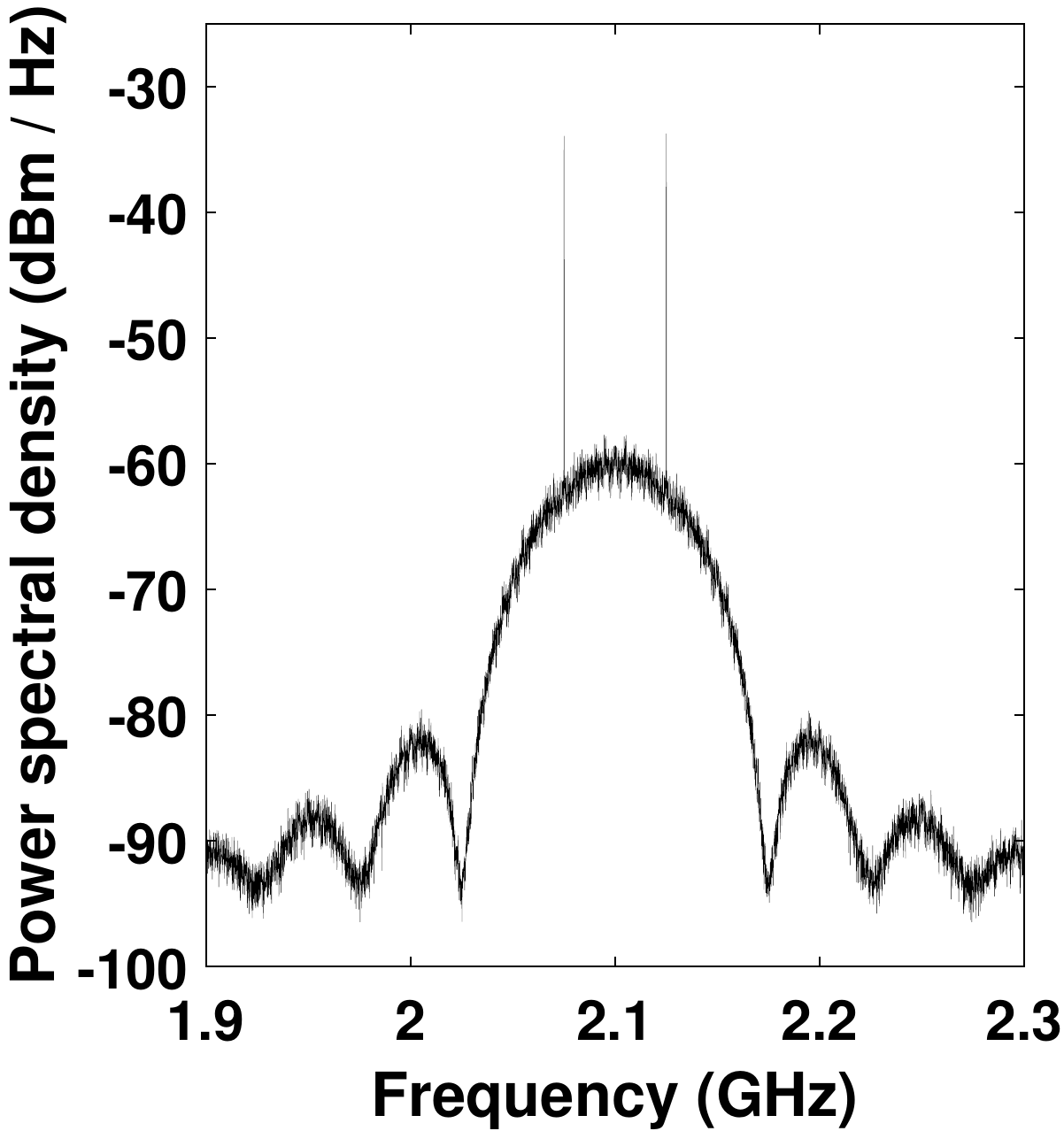}}
     \hspace{0.005cm}
     \subfloat[Nonlinear PA\label{subfig-2:FSK2}]{%
       \includegraphics[width=0.22\textwidth, height = 0.15\textwidth]{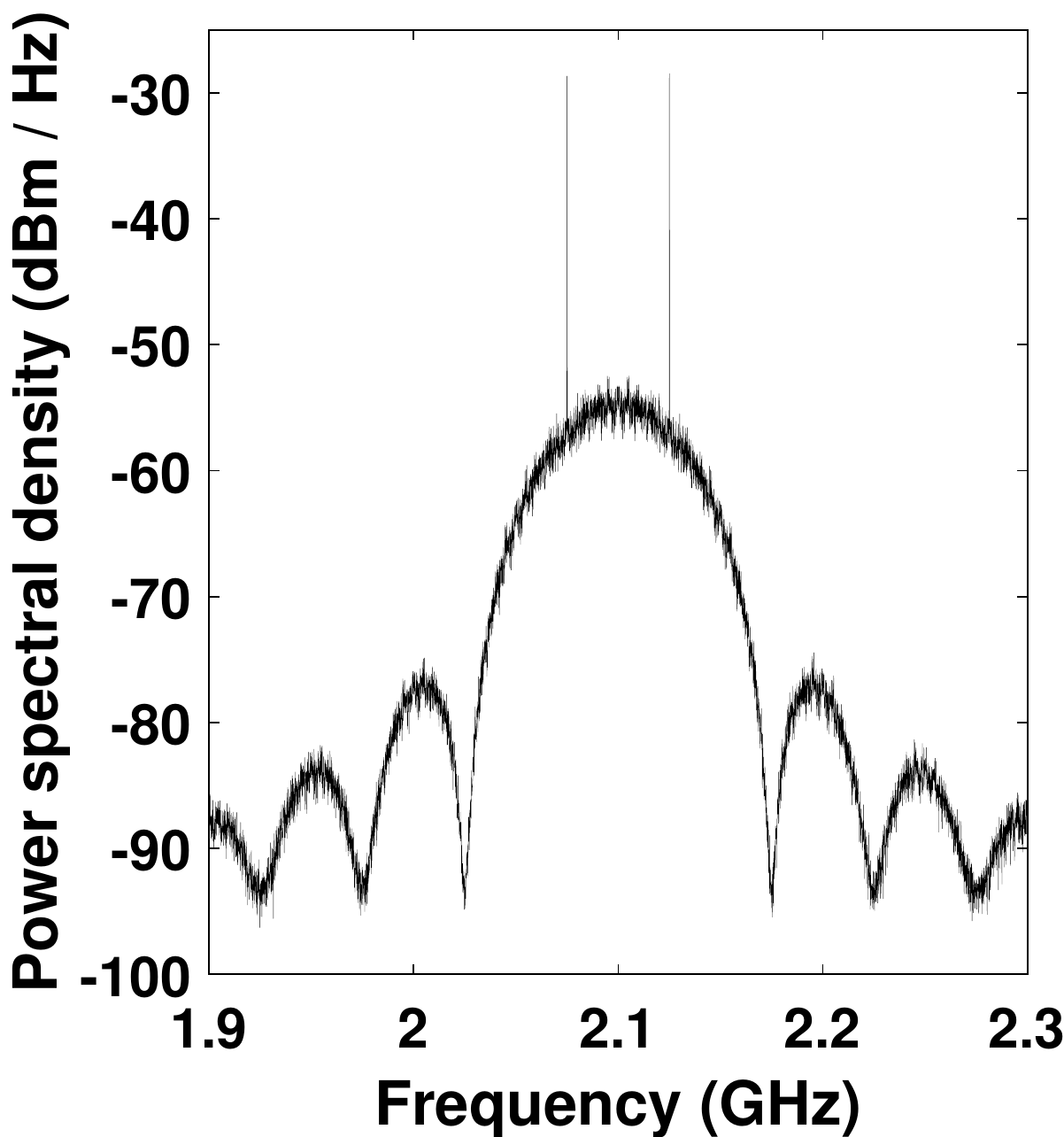}}
       \caption{Nonlinearity effect under BFSK modulation}
     \label{fig:FSK_PA}
   \end{figure}

\begin{figure}
     \subfloat[Linear PA\label{subfig-1:Lin}]{%
       \includegraphics[width=0.152\textwidth]{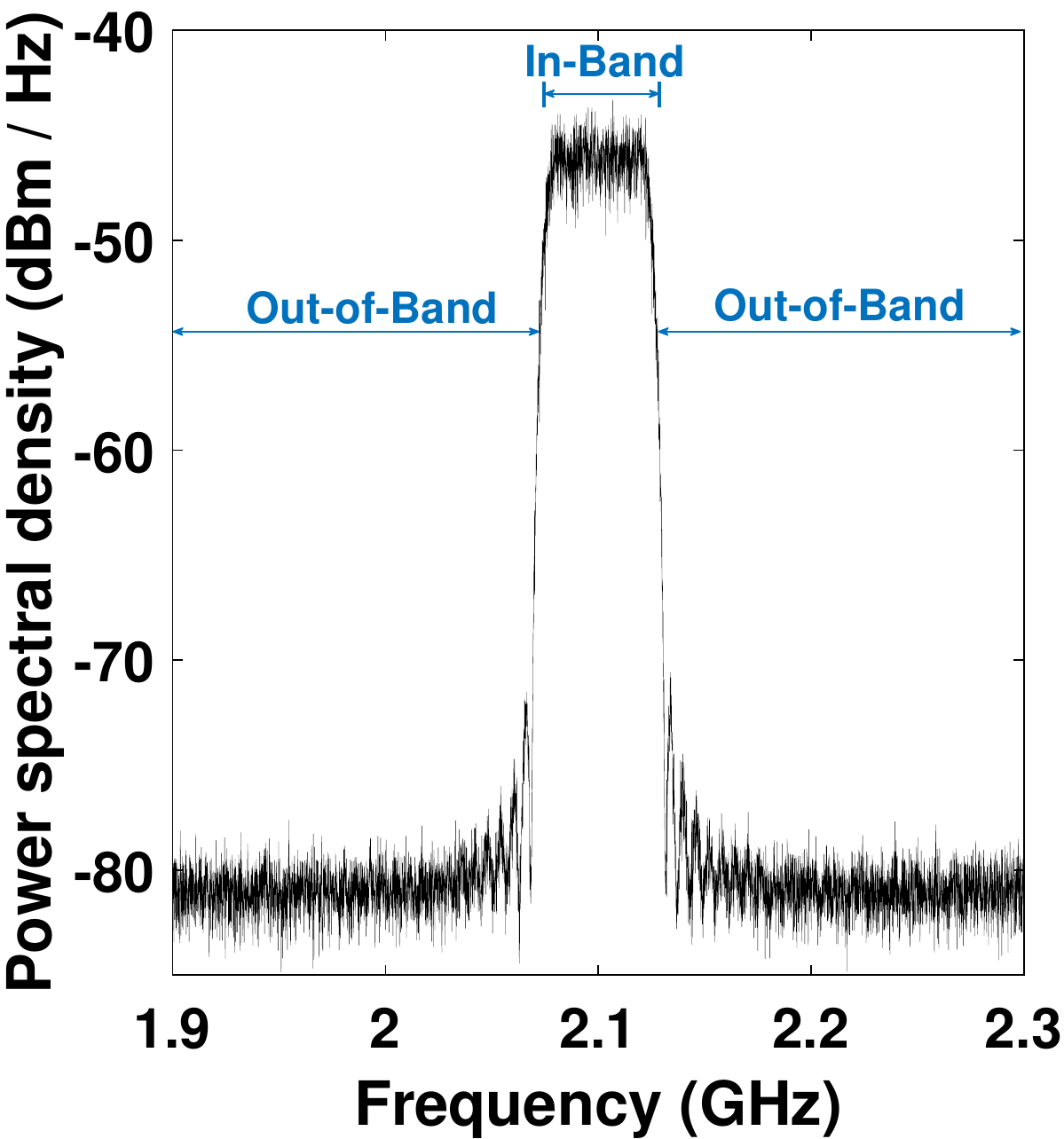}
     }
     \hspace{0.005cm}
     \subfloat[Nonlinear PA 1\label{subfig-2:Non1}]{%
       \includegraphics[width=0.152\textwidth]{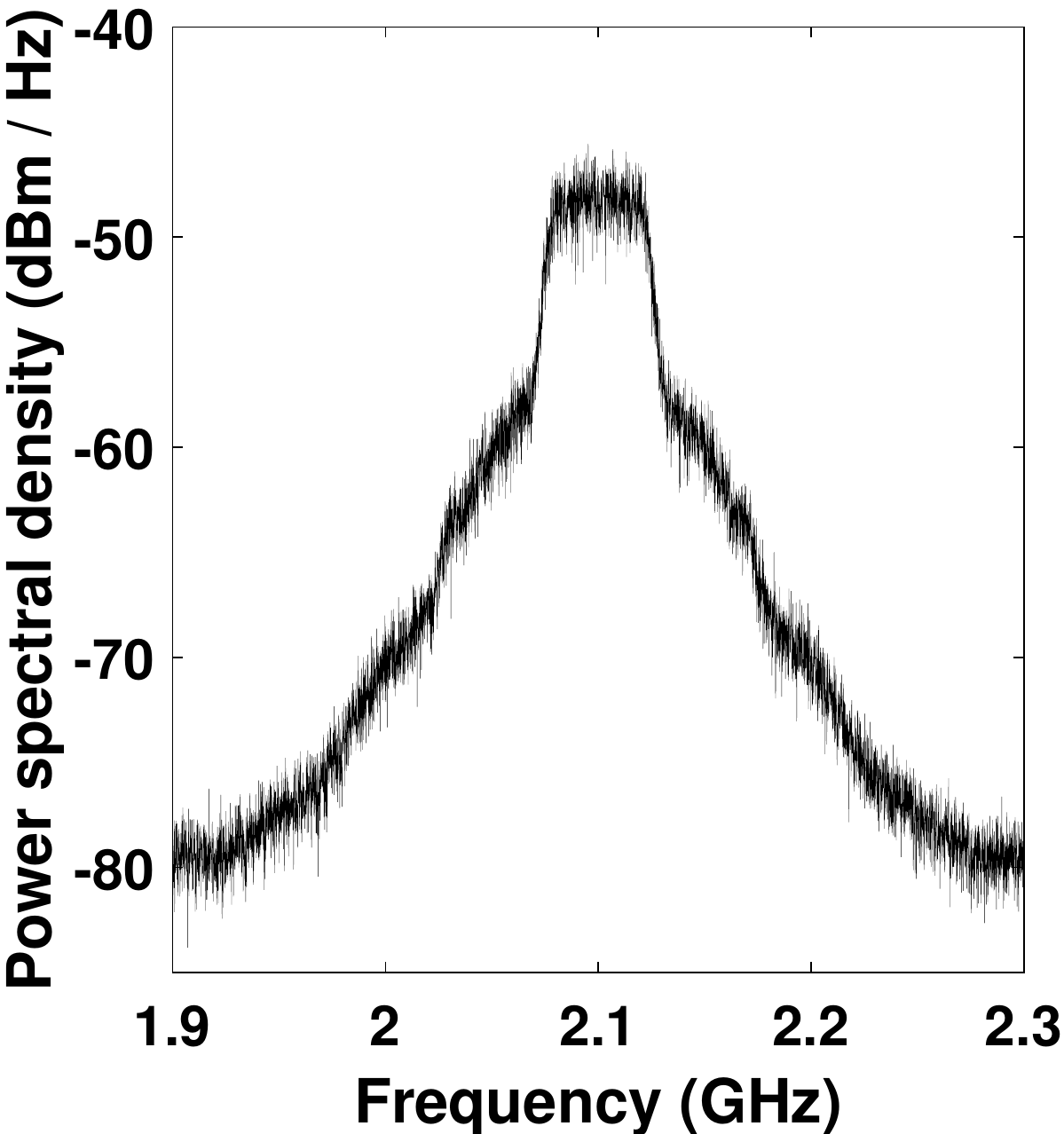}
     }
     \hspace{0.005cm}
     \subfloat[Nonlinear PA 2\label{subfig-2:Non2}]{%
       \includegraphics[width=0.152\textwidth]{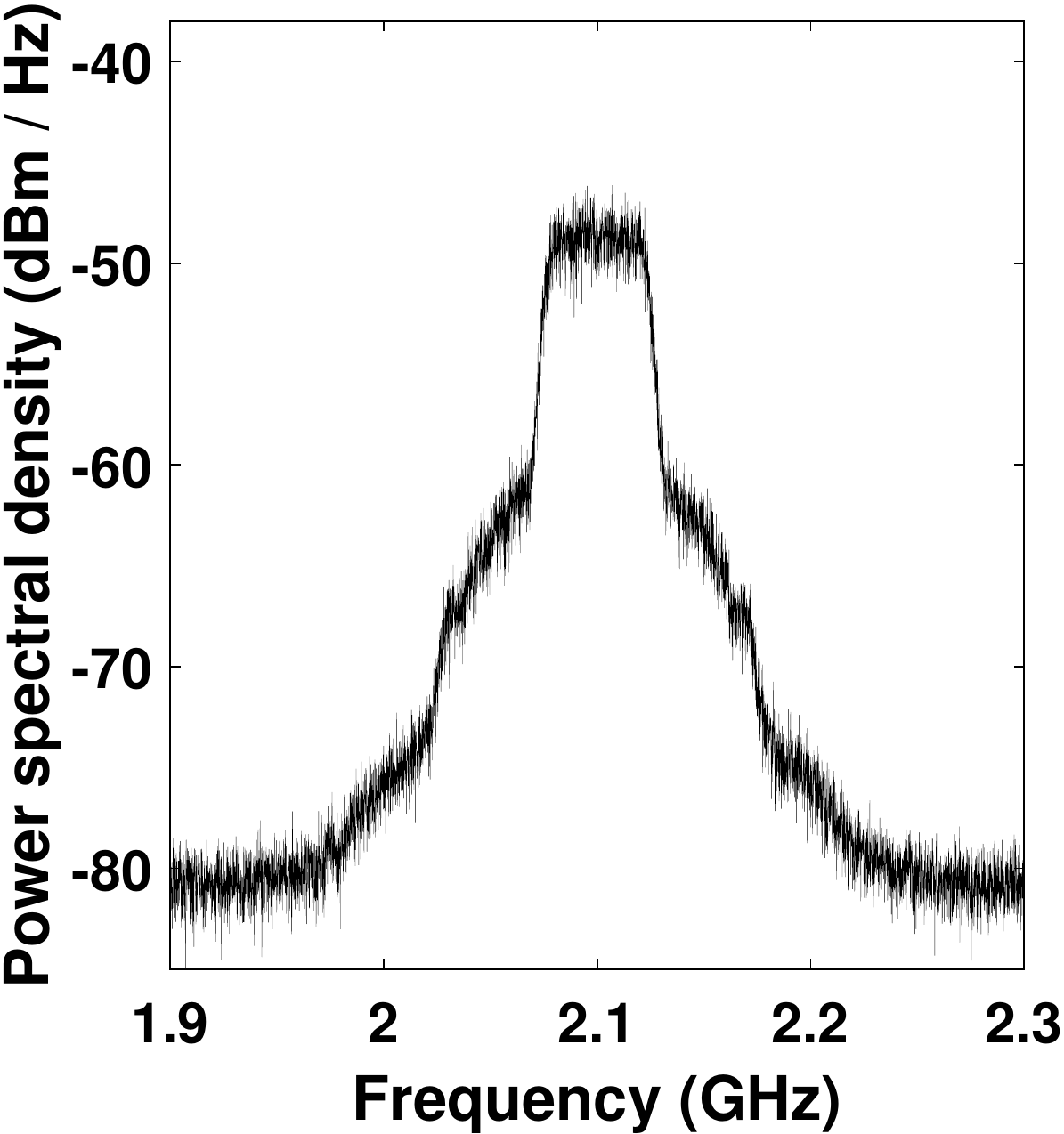}
     }
     \caption{Nonlinearity effect under 16QAM modulation}
     \label{fig:PA}
   \end{figure}

Recall that Eq.~\eqref{eq:5} expresses the output signal of nonlinear/real PA as a function of all odd nonlinear terms. 
For ease of illustration, let's look at the effect of the third-order nonlinearity term only, when feeding the output signal, $S_{RF}(t)\!\!= \!\!A(t)\cos(w_ct+\phi(t))$, of the in-phase branch mixer as an input to the PA. In this case, the PA output is  
${S}_{PA}(t)=\Tilde{\alpha}_1S_{RF}(t)+\Tilde{\alpha}_3S_{RF}^3(t)$, with the third-order nonlinearity term,  $\Tilde{\alpha}_3S_{RF}^3(t)$, being
%
%
\begin{equation}\nonumber
    \Tilde{\alpha}_3S_{RF}^3\!(t)\!=\!\frac{\Tilde{\alpha}_3A^3(t)}{4}\big[3\cos(w_ct+\phi(t)) + \cos(3w_ct+3\phi(t))\big]
\end{equation}
where $\Tilde{\alpha_1}$ and $\Tilde{\alpha_3}$ are again the complex coefficients modeling the nonlinearity terms. Note that given that the out-of-band component at $3w_c$ is located sufficiently far away from the center frequency, $w_c$, and that the bandwidth of the original signal is much less than $w_c$, this out-of-band component can easily be filtered out without causing any bandwidth regrowth around the original signal spectrum. However, the first term at $w_c$ may lead to spectrum regrowth of the original message bandwidth, depending, for example, on the modulation technique being used.
For instance, in the case of constant-envelope modulation schemes such as BPSK where the amplitude $A(t)$ is constant, the spectrum of the modulated signal in the vicinity of $w_c$ remains unchanged. This can be shown in Fig.~\ref{fig:FSK_PA} where the spectrum of a BFSK modulated signal has not changed after passing through a nonlinear PA. Note that the shape of the spectrum is the same under both linear and nonlinear PAs.
However, for variable-envelope modulation schemes such as $16$QAM where the amplitude $A(t)$ varies over time, because the ${\Tilde{\alpha}_3^3A^3(t)}/{4}$ term generally exhibits a broader spectrum than $A(t)$ itself, 
nonlinearity causes spectral regrowth. 
%
For this case of modulation, the severity of the spectral growth also depends on the nonlinearity model parameter $\Tilde{\alpha}_3$. 
To illustrate, we show in Fig.~\ref{fig:PA} the case of a 16QAM modulated signal passing through a linear PA (Fig.~\ref{subfig-1:Lin}) and two nonlinear PAs (Figs.~\ref{subfig-2:Non1} and~\ref{subfig-2:Non2}) each under slightly different nonlinearity parameters.
Two key observations we make from these results. First, the nonlinearity of PA does lead to an out-of-band spectrum growth (or distortion). Second, even a slight difference in the nonlinearity impairments causes differences in the amplitude of the frequency components in the out-of-band domain, as one can observe from the indicated amplitudes of the spikes. That is, even a slight nonlinearity impairment difference causes quite different out-of-band spectrum distortions. Our proposed classification technique exploits this out-of-band distortion information to increase both the accuracy and scalability of device classification.



\begin{figure}
    \centering
    \includegraphics[width=6cm]{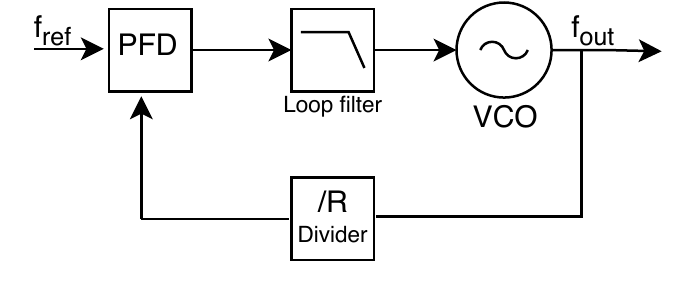}
    \caption{A simple PLL schematic}
    \label{fig:pll}
\end{figure}

\begin{figure}
     \subfloat[Device 1\label{subfig-1:ph1}]{%
       \includegraphics[width=0.157\textwidth, height = 0.161\textwidth]{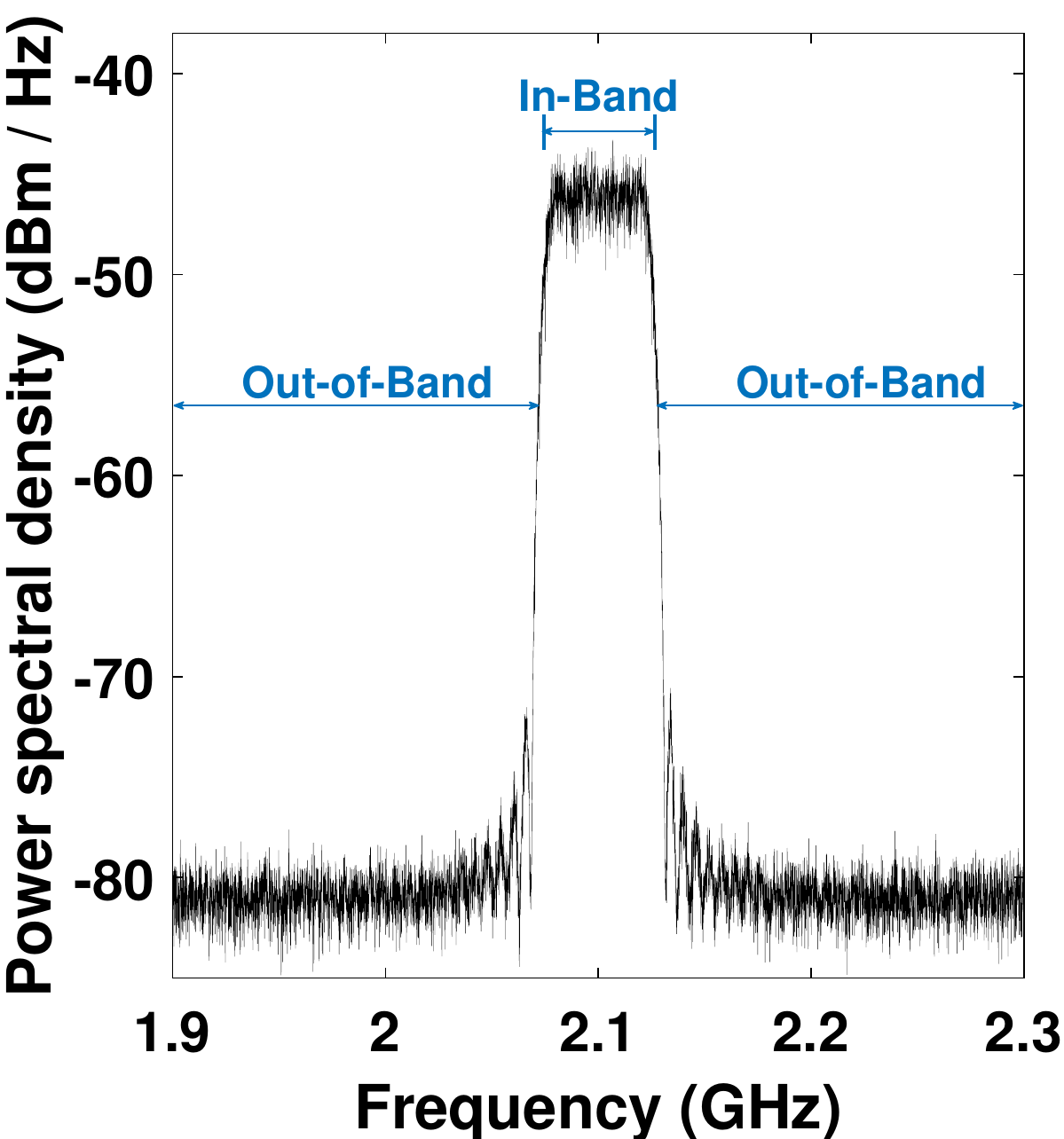}}
     \hspace{0.005cm}
     \subfloat[Device 2\label{subfig-2:ph2}]{%
       \includegraphics[width=0.157\textwidth, height = 0.164\textwidth]{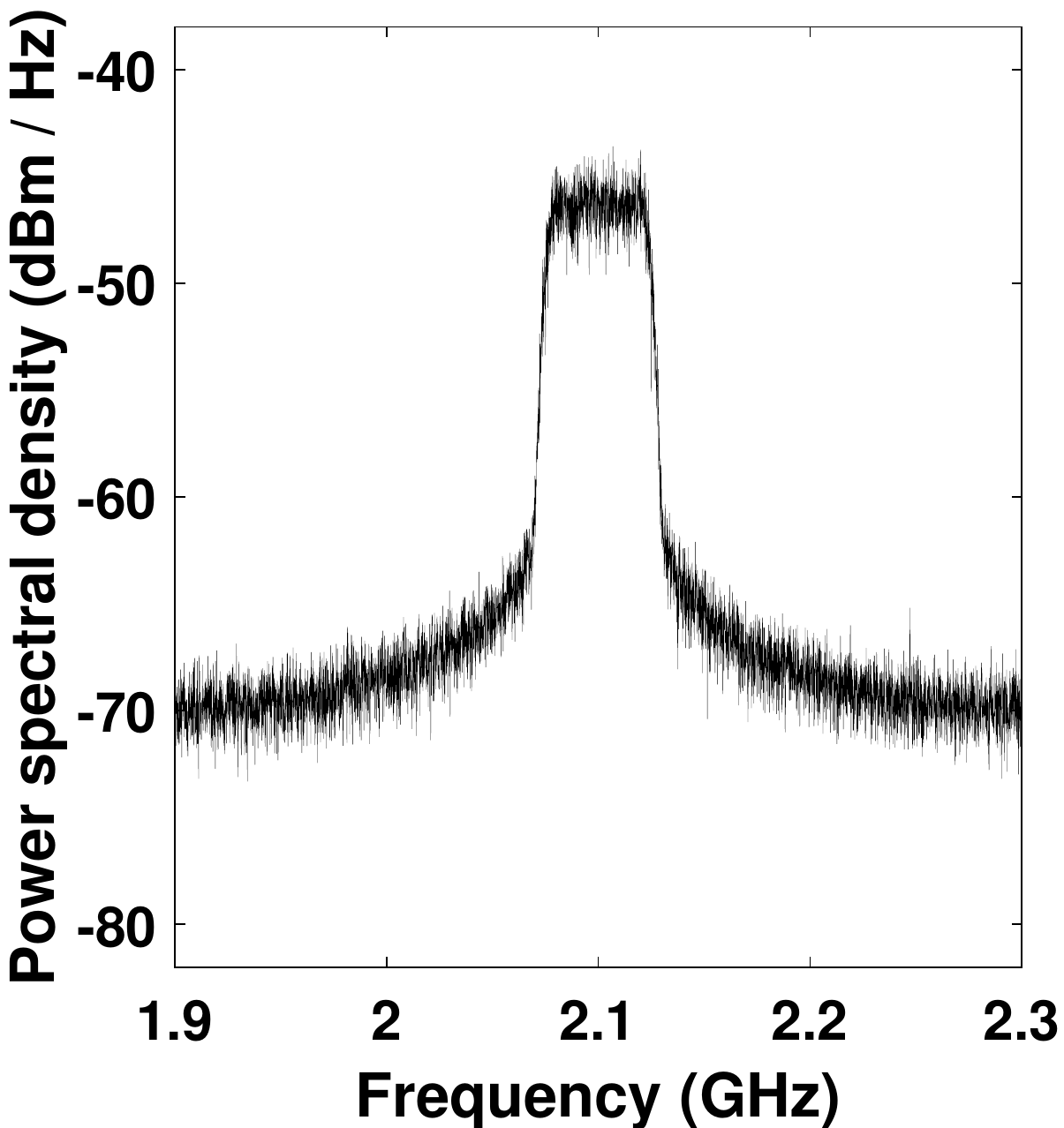}}
     \hspace{0.005cm}
     \subfloat[Device 3\label{subfig-2:ph3}]{%
       \includegraphics[width=0.157\textwidth, height = 0.164\textwidth]{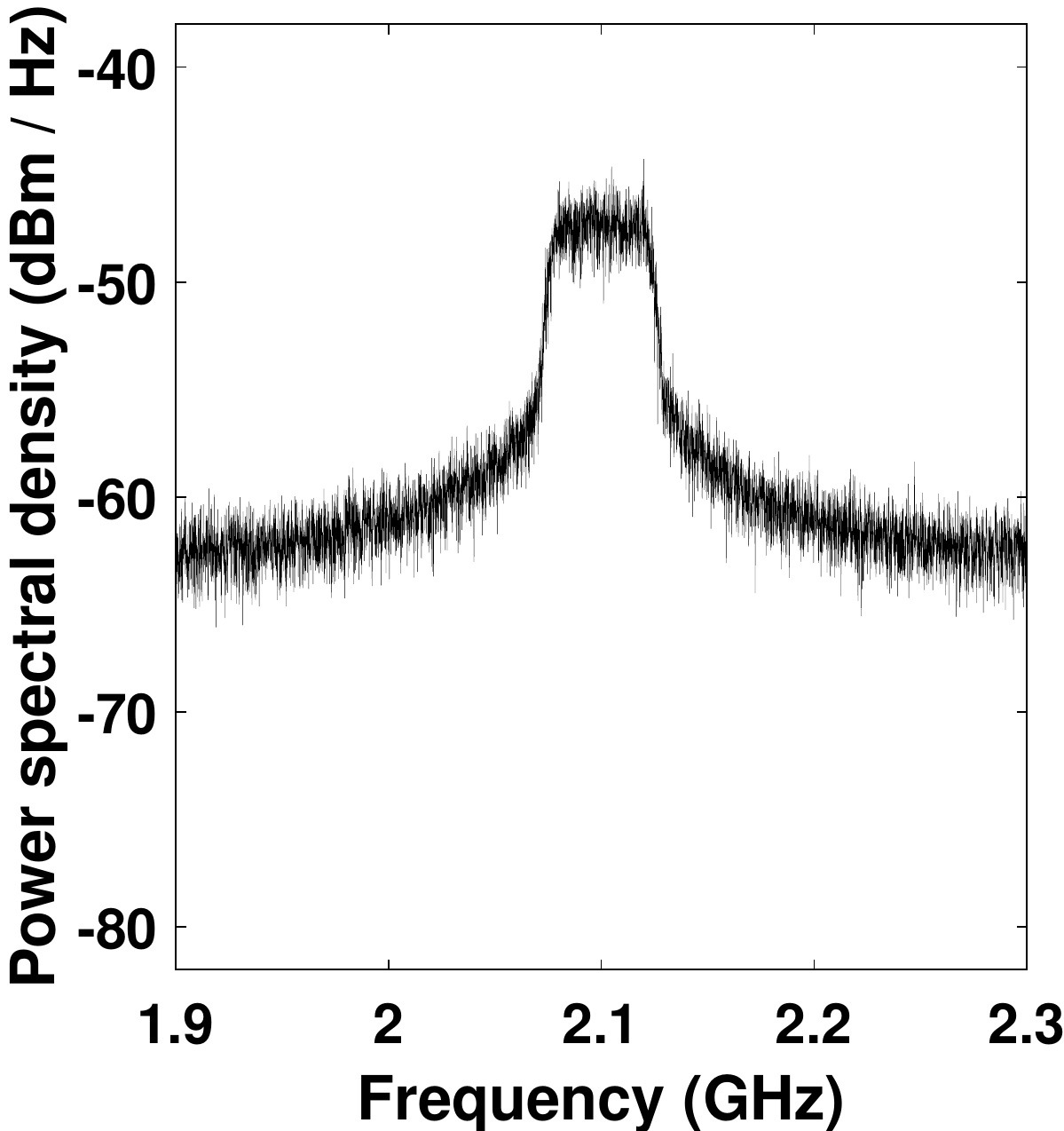}
     }
     \caption{Phase Noise effect: Device 1 (ideal); Device 2 (-80 dBc/Hz); Device 3 (-72 dBc/Hz); at the frequency offset $1$MHz.}
     \label{fig:Phase_noise}
   \end{figure}

\subsubsection{Local Oscillator (LO)}
In modern transceivers, LOs are usually made with Phase-Locked Loops (PLLs) that ensure high-frequency stability and minimum phase noise~\cite{smaini2012rf}. Fig.~\ref{fig:pll} shows a simple schematic of a PLL where $f_{ref}$ is the reference frequency, $f_{out}$ is the output frequency, PDF is the phase-frequency discriminator, and the VCO is the voltage-controlled oscillator. Looking into the transfer functions of a PLL's components, which compose the closed-loop transfer function of the PLL, provides an insight into the noise contribution of each of them. The transfer function of PLL from the reference frequency to VCO, for example, has a low-pass characteristic and can be expressed as~\cite{gardner1979phaselock} $ H_{ref}(s) = {R(2\epsilon w_ns + w_n^2)}/{(s^2 + 2\epsilon w_ns + w_n^2)}$
where $R$ is the feedback divider and $w_n$ and $\epsilon$ are the natural frequency and the damping coefficient, respectively. 
The transfer function of the VCO, on the other hand, has a high-pass characteristic and can be defined as $H_{VCO}(s) = {s^2}/{(s^2 + 2\epsilon w_ns + w_n^2)}$.
%
Hence, PLL in-band phase noise is dominated by the three components described above that have low-pass characteristics, while the out-of-band noise is mainly a function of the impaired VCO~\cite{smaini2012rf}. 

To illustrate the impact of phase noise on out-of-band distortion, consider the mixer output signal in the in-band path, $S_{I_{RF}}(t) = S_I(t)\cos(w_ct+ \theta(t))$ as given in Eq.~\eqref{eq:7}, where $\theta(t)$ is again the LO phase noise. Applying the Fourier transform to both sides of this mixer output equation yields
\begin{equation}\label{eq:mxer}
     \mathcal{F}[S_{I_{RF}} \!(t)]\!\!=\!\!  \frac{1}{2}\big\{\bar{S}_I(f\!-\!f_c)*\mathcal{F}[e^{j\theta(t)}]\!+\!\bar{S}_I\!(f\!+\!f_c)*\mathcal{F}[e^{-j\theta(t)}]\big\}
\end{equation}
where $f_c=2\pi w_c$, $\bar{S}_I(f)=\mathcal{F}[S_I(t)]$, and $\mathcal{F}[.]$ and * are the Fourier transform and convolution operators. Eq.~\eqref{eq:mxer} shows that there is a bandwidth expansion around the carrier frequency $f_c$ beyond the spectrum of the original signal, resulting from the convolution of the original signal spectrum and the spectrum of LO impairment term $e^{-j\theta(t)}$. 

Now since the spectrum expansion (or regrowth) is a function of the LO phase noise term, $e^{\theta(t)}$, different devices will exhibit different spectral regrowth; i.e., different out-of-band distortions. 
This can be clearly seen in Fig.~\ref{fig:Phase_noise}, where the PSD of three simulated devices, each with different phase noise value, but at the same frequency offset, are displayed. Device $1$ enjoys an ideal LO (i.e., zero phase noise value), while device $2$ and device $3$ suffer from a phase noise value of $-80$ and $-72$ dBc/Hz, respectively, at the same frequency offset, 1MHz. Therefore, considering the out-of-band information makes the spectra of devices more discernible and thus enhances the performance of the classifier. Our proposed classification technique exploits out-of-band distortion information caused by LO phase noise to improve classification accuracy and device separability.

\subsubsection{Mixers}
\begin{figure}
     \subfloat[Device 1\label{subfig-1:dc1}]{%
       \includegraphics[width=0.157\textwidth, height = 0.168\textwidth]{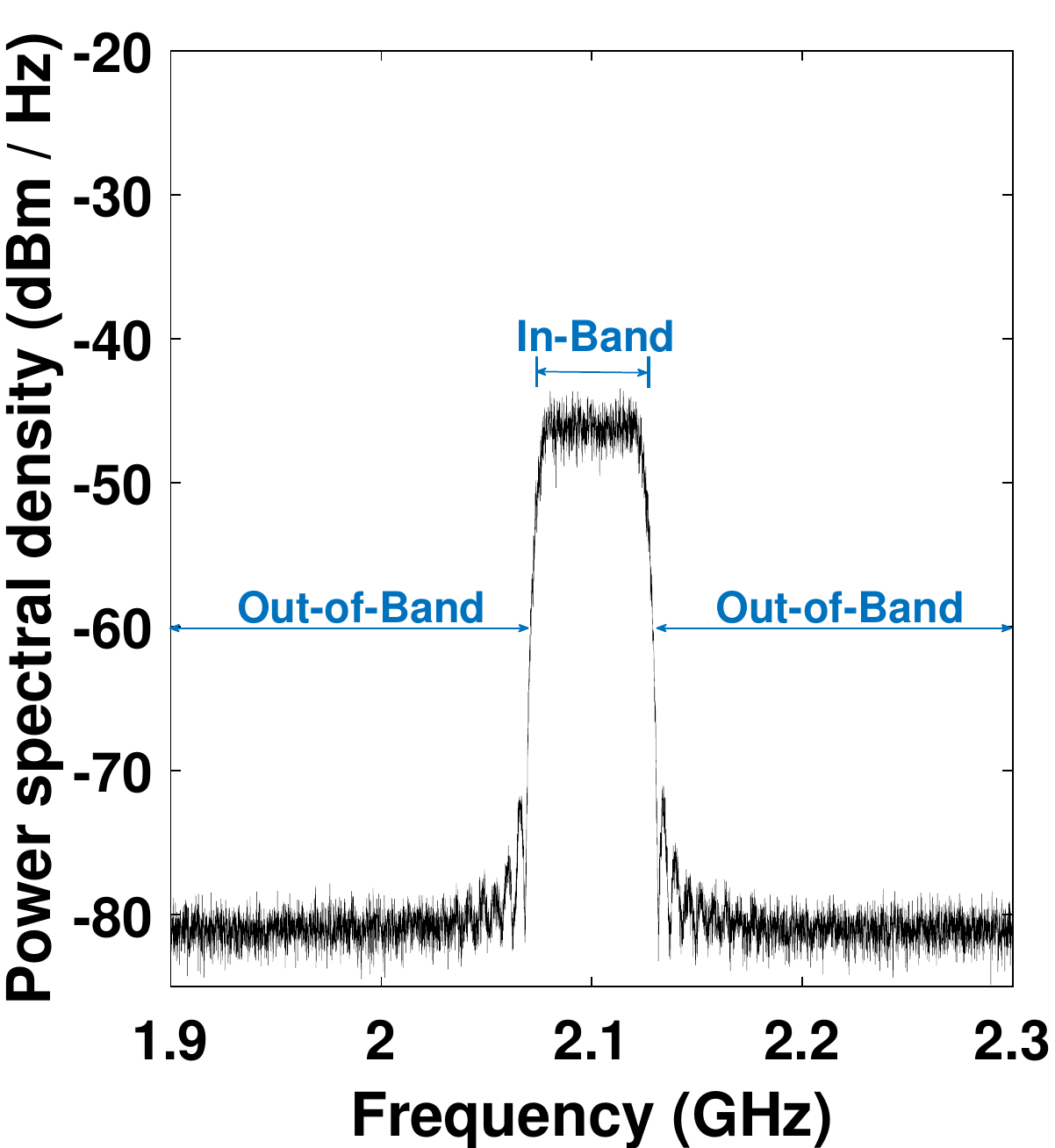}}
     \hspace{0.005cm}
     \subfloat[Device 2\label{subfig-2:dc2}]{%
       \includegraphics[width=0.157\textwidth, height = 0.168\textwidth]{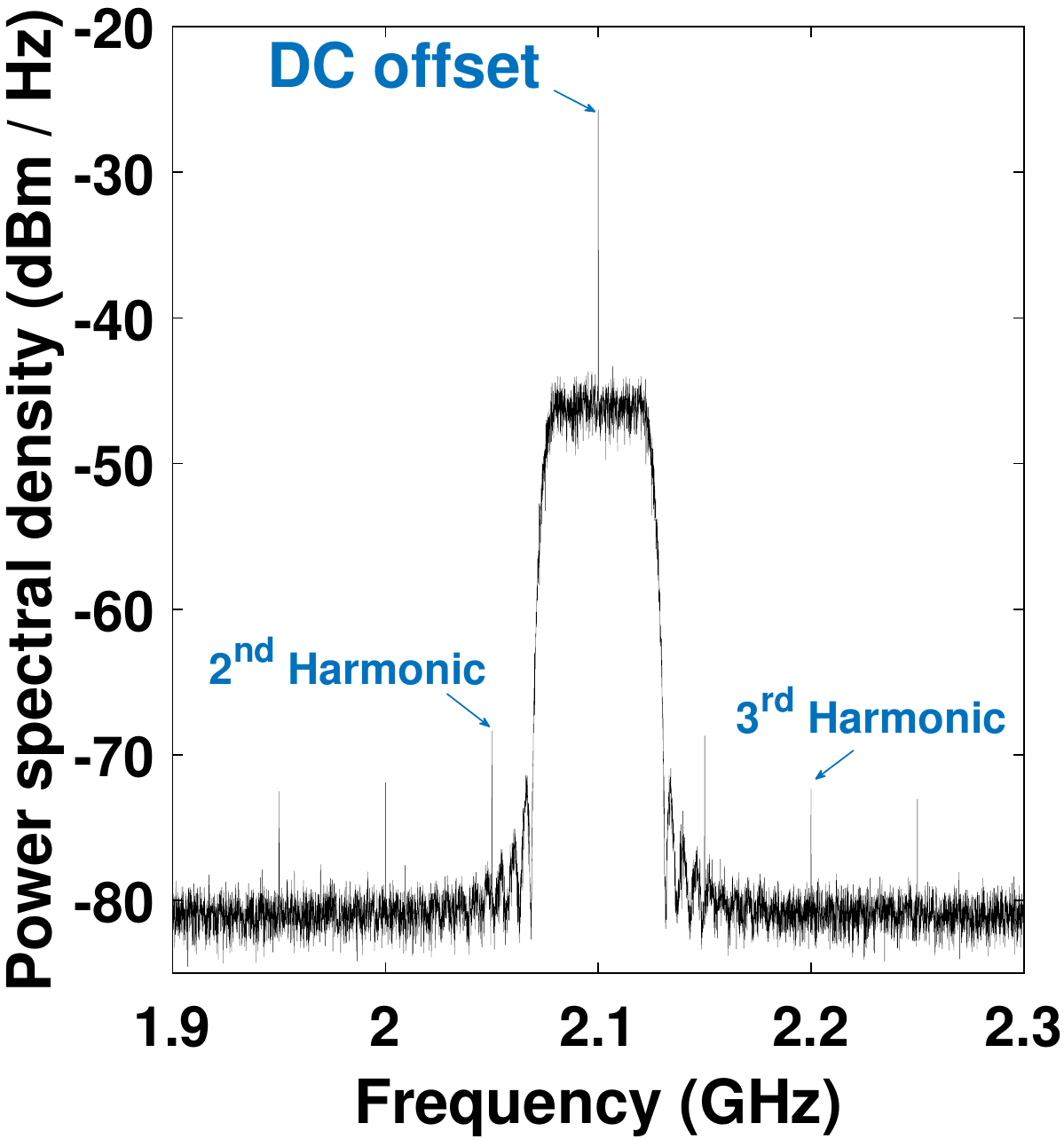}}
     \hspace{0.005cm}
     \subfloat[Device 3\label{subfig-2:dc3}]{%
       \includegraphics[width=0.157\textwidth, height = 0.168\textwidth]{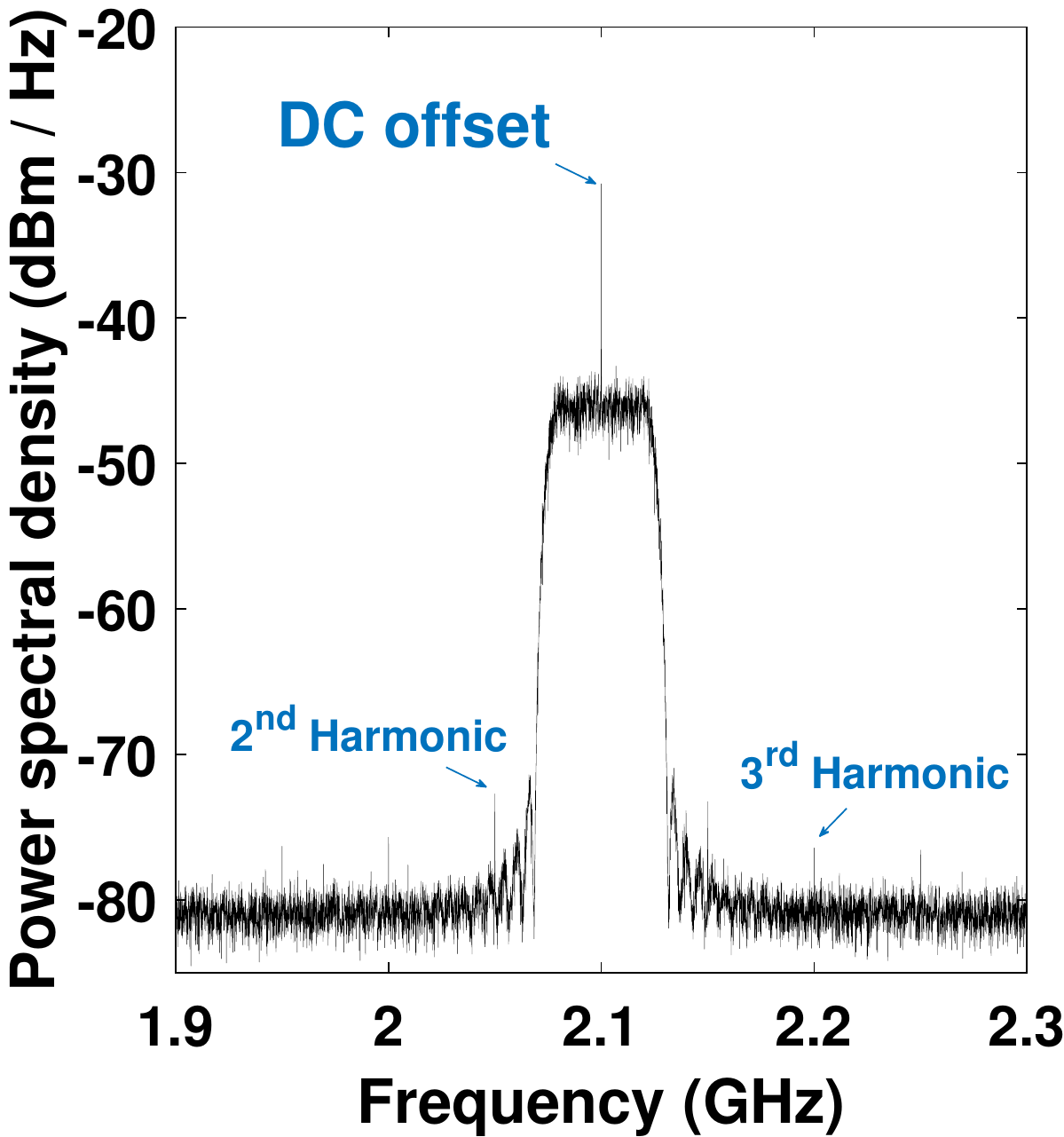}
     }
     \caption{DC Offset Effect: Device $1$ (ideal mixer, DC offset = $0$); Device $2$ (DC offset: I =  $0.9$ and Q = $0.9$); Device $3$ (DC offset: I = $0.5$ and Q = $0.5$)}
     \label{fig:dc-offset}
   \end{figure}

Beside the relative large DC component at the center of the signal spectrum that real mixes introduce, the nonlinearity of the mixer also introduces other undesired harmonic spurs within the out-of-band domain. The amplitude of the DC component and its harmonics depend on both the silicon-level circuitry of the mixer and the second-order nonlinearity distortion of the device. This can be clearly observed by comparing the amplitudes of the spikes shown in the PSD of the three simulated devices in  Fig.~\ref{fig:dc-offset}. Device $1$ mimics an ideal mixer (i.e., zero DC offset), while device $2$ and device $3$ mimic real mixers with in-phase DC offset values of $0.9$ and $0.5$ and quadrature offset values of $0.9$ and $0.5$, respectively. Observe, from the figure, that ideal mixers do not yield any DC component nor its harmonics, whereas hardware-impaired, real mixers yield DC spurs at the center of the spectrum as well as in the out-of-band region. Also, observe that the amplitudes of the DC spurs of device $2$ and device $3$  occurring in both the in-band and the out-of-band spectrum are different from one another, even though the differences between their DC offset values are insignificant. Therefore, a transmitter's DC component and its harmonic spurs caused by mixer impairments can potentially be leveraged for providing unique device signature that can be used for device classification. 
Our proposed classification technique leverages the out-of-band information that captures the differences between the DC offset harmonic spurs of devices to increase device separability classification accuracy.  

\subsubsection{Digital to Analog Converter (DAC)}
DACs also suffer from nonlinearities and hardware impairments that can be exploited to provide unique features and signatures for devices. In addition to degrading the error vector magnitude (EVM) of a transmitter, DAC impairments are responsible for generating out-of-band (OOB) emissions as well. To quantify and illustrate these OOB distortions, we refer to Eq.~\eqref{eq8}, which models the DAC output when considering the in-phase signal component, $S_I[n]$, as its input, while capturing the three main distortions, horizontal quantization (HQ), vertical quantization (VQ), and clock source modulation (CM), caused by the DAC. Note that although here we focus on the in-phase (I) path component for illustration purposes, similar analysis and illustration can be done for the case of the quadrature (Q) path component.
%
%
Even though each of the three DAC impairments, HQ, VQ, and CM, yields OOB emissions, HQ contributes the most when the DAC generation frequency is not sufficiently greater than the Nyquist rate, and hence, we focus only on HQ's impact in this illustration. 
Using Fourier series representations, the HQ term, $y_{S_I}^{HQ}(t)$, can be written as $y_{S_I}^{\rm HQ}(t) = \sum_{k = -\infty}^{\infty}c_{k}^{\rm HQ}e^{j2 \pi kf_{\Omega}t}$,
with the Fourier coefficients $c_{k}^{\rm HQ}$ being
\begin{equation}
    c_{k}^{\rm HQ} = {\frac{1}{ T_{\Omega}}}\!\!\int\limits_{0}^{T_{\Omega}}\!\!\big(\!\!\!\sum_{n = -\infty}^{\infty}\!\!S_I[n]g\left(\frac{t - nT_{g}}{ T_{g}}\right) - S_I(t)\big)e^{-j2\pi kf_{\Omega}t}\,dt
    \label{eq:13}
\end{equation}
where $T_{\Omega}$ is time period of the three distortion additive terms in Eq.~\eqref{eq8}, which is the least common period of the three periods: output signal period, $T_0$, DAC generation period, $T_g$, and the clock modulation period, $T_m$ \cite{d2010modeling}. Leveraging the fact that $g(\theta)$ is a unitary pulse only when $0\!\leq\!\theta\!<\!1$ and $T_{\Omega}\!\!=\!\!ZT_g$, we can extend the integral to $(-\infty, \infty)$, while restricting the index $\!n\!$ in the first term from $0$ to $Z-1$ and introducing a unitary window in the second term $W_{[0, T_{\Omega}]}$. Then, we can rewrite Eq.~\eqref{eq:13} as~\cite{d2010modeling}:
\begin{equation}
    c_{k}^{\rm HQ}\!=\!{1\over T_{\Omega}}\!\!\!\int\limits_{-\infty}^{{\infty}}\!\!\!\big(\!\sum_{n = 0}^{Z-1}\!\!S_I[n]g\!\left(\!\!{t - nT_{g}\over T_{g}}\!\!\right)\!-\! S_I(t)W_{[0,T_{\Omega}]}\big)e^{-j2\pi kf_{\Omega}t}\!dt \nonumber
    \label{eq:14}
\end{equation}
By Fourier-analyzing the second term in the right-hand side of the above equation, we observe that 
the spectral contribution of the second term would be samples of the spectrum of the distorted version of $S_I(t)$ at frequencies $kf_{\Omega}$, with $k$ ranging from $0$ to $Z-1$, which lie mostly outside the bandwidth of $S_I(t)$. 
%
Therefore, most effects of the $y_{S_I}^{HQ(t)}$ term lie outside the bandwidth of $S_I(t)$~\cite{d2010modeling}, resulting in the growth of the number of attenuated replicas in the out-of-band domain of the signal $S_I(t)$.


\section{Performance Evaluation and Analysis}
\label{Performance_eval}
\begin{figure*}
    \centering
    \includegraphics[width=1.7\columnwidth]{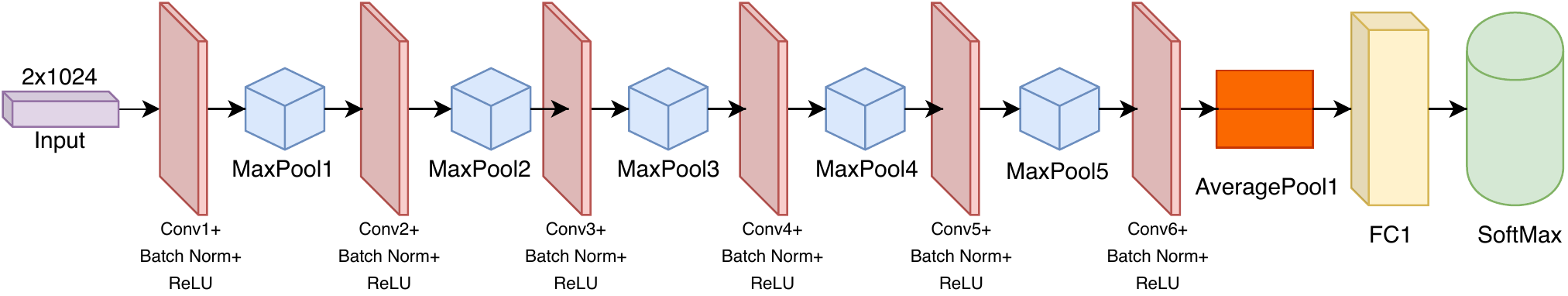}
    \caption{CNN architecture used in our experiment}
    \label{fig:cnn}
\end{figure*}

\begin{table*}
\vspace{0.1in}
\centering

\resizebox{16cm}{!}{

\begin{tabular}{|l|l|l|l|l|l|l|l|l|}
\hline
              &                     &                         &                      &                      &                                           &                     &                               &                          \\ 
\textbf{RF}   & \textbf{IQ-amp(dB)} & \textbf{IQ-phase(deg)} & \textbf{I-DC offset} & \textbf{Q-DC offset} & \multicolumn{1}{c|}{\textbf{AM-AM}}       & \textbf{AM-PM}      & \textbf{Phase noise(dBc/Hz)} & \textbf{Freq offset(Hz)} \\ \hline
              &                     &                         &                      &                      &                                           &                     &                               &                          \\ 
\textbf{Dev1} & 0.08                & 0.1                     & 0.1                  & 0.15                 & \multicolumn{1}{c|}{{[}2.178 1.12157{]}}  & {[}4.0893 9.2040{]} & {[}-60, -80{]}                & {[}20, 200{]}            \\ \hline
              &                     &                         &                      &                      &                                           &                     &                               &                          \\ 
\textbf{Dev2} & 0.1                 & 0.09                    & 0.109                & 0.1                  & \multicolumn{1}{c|}{{[}2.197 1.16157{]}}  & {[}4.13 9.2540{]}   & {[}-60, -80{]}                & {[}20, 200{]}            \\ \hline
              &                     &                         &                      &                      &                                           &                     &                               &                          \\ 
\textbf{Dev3} & 0.09                & 0.09                    & 0.1                  & 0.1                  & \multicolumn{1}{c|}{{[}2.16 1.10157{]}}   & {[}4.0933 9.2840{]} & {[}-59.9, -80{]}              & {[}20, 200.9{]}          \\ \hline
              &                     &                         &                      &                      &                                           &                     &                               &                          \\ 
\textbf{Dev4} & 0.109               & 0.108                   & 0.1                  & 0.1                  & \multicolumn{1}{c|}{{[}2.17 1.12157{]}}   & {[}4.113 9.2040{]}  & {[}-60, -80.1{]}              & {[}20, 200{]}            \\ \hline
              &                     &                         &                      &                      &                                           &                     &                               &                          \\ 
\textbf{Dev5} & 0.1                 & 0.099                   & 0.099                & 0.1                  & \multicolumn{1}{c|}{{[}2.1587 1.15157{]}} & {[}4.133 9.2040{]}  & {[}-60, -80{]}              & {[}20.1, 200{]}         \\ \hline
\end{tabular}

}
\caption{Transceiver hardware impairments}
    \label{fig:table1}
\end{table*}

MATLAB's Communications toolbox is used to design a simulation model of full wireless communications processing chain for $5$ different devices. Each device represents a transmitter that sends $16$QAM modulated signals over an AWGN channel. 
Different RF impairments blocks are used to introduce and set different values for IQ imbalance, DC offset, carrier frequency offset, phase noise, and PA nonlinearity distortion. Table~\ref{fig:table1} shows the different impairment values used in our experiment.
IQ imbalance values are shown in the first and the second columns of the table, where the first one represents the amplitude mismatch, IQ-amp, and the second column represents the phase deviation, IQ-phase. The in-phase DC offset and the quadrature DC offset values are presented in the third and fourth columns. The PA nonlinearity distortion is represented in the fifth and sixth columns by the alpha and beta parameters of Saleh model functions of the Amplitude-to-Amplitude (AM-AM)  and  Amplitude-to-Phase (AM-PM) distortion curves~\cite{saleh1981frequency}. The last two columns of the table show the LO phase noise introduced by a filtered Gaussian noise using a spectral mask specified by noise level and the frequency offset vectors.

For each device, we collect the raw IQ values of two different bandwidths, $2.075$ - $2.125$ GHz, which represents the bandwidth of the message (in-band), and $1.9$ - $2.3$ GHz, which includes both in-band (message bandwidth) and out-of-band domain. We generate $200$k samples for each  device, which are divided into training, validation, and test sets.


\subsection{CNN Classifier Architecture}
We design a CNN architecture that uses raw time-series IQ samples generated by our Simulink model.
We use a variation of the CNN architecture used in~\cite{liu2017deep}, which is depicted in Fig.~\ref{fig:cnn}.
Specifically, each IQ input sequence is represented as a two dimensional real-valued tensor of size $2$×$1024$. Thus, the in-phase (I) and quadrature (Q) components are processed independently and only in the fully connected layer where the information of the two components combined. The input is fed to
the first convolutional layer (Conv1), which consists of $16$ filters, each of size $1$x$4$. Each filter learns $4$-sample variations in time over
the I or Q dimension separately to generate $16$ distinct feature maps
over the complete input sample. Each ConvLayer is followed by a Batch normalization layer, a
Rectified Linear Unit (ReLU) activation, and a maximum pooling
(MaxPool) layer with filters of size $1$x$2$ and stride [$1$ $2$] to perform a
pre-determined non-linear transformation on each element of the
convolved output, except the last ConvLayer, which is followed by an Average Pooling (AP) layer with a dimension $1$x$32$. 
The output of the AP layer is then provided
as an input to the Fully Connected (FC) layer, which has $5$ neurons. Then, the output of the FC is finally passed to a classifier layer. To
overcome overfitting, we set the dropout rate to $0.5$ 
at the dense layers. A softmax classifier is used in the last layer to output the
probabilities of each frame being fed to the CNN.

Weights are trained using stochastic gradient descent with momentum optimizer with an initial learning rate of $l=0.02$ and a learning rate drop factor of $0.1$ with a learning rate drop period of $9$. We minimize the prediction error through
back-propagation, using categorical cross-entropy as a loss function computed on the classifier output. We implement our CNN architecture in MATLAB using the Deep Learning Toolbox running on a system
with intel Corei7 8th Gen CPU.

\subsection{Result Analysis}
We evaluate and compare the performance of the proposed technique, leveraging both in-band and out-of-band spectrum distortions, and the conventional classification technique, using in-band distortion information only. The impairments values used in our experiment, which are shown in Table~\ref{fig:table1}, are set very similar to one another so that the devices resemble bit-similar radios to make the identification task even harder.

\begin{figure}
	\centering
	\subfloat[Our proposed in-band and out-of-band technique.]{
    \includegraphics[width=.85\columnwidth,height=.18\textheight]{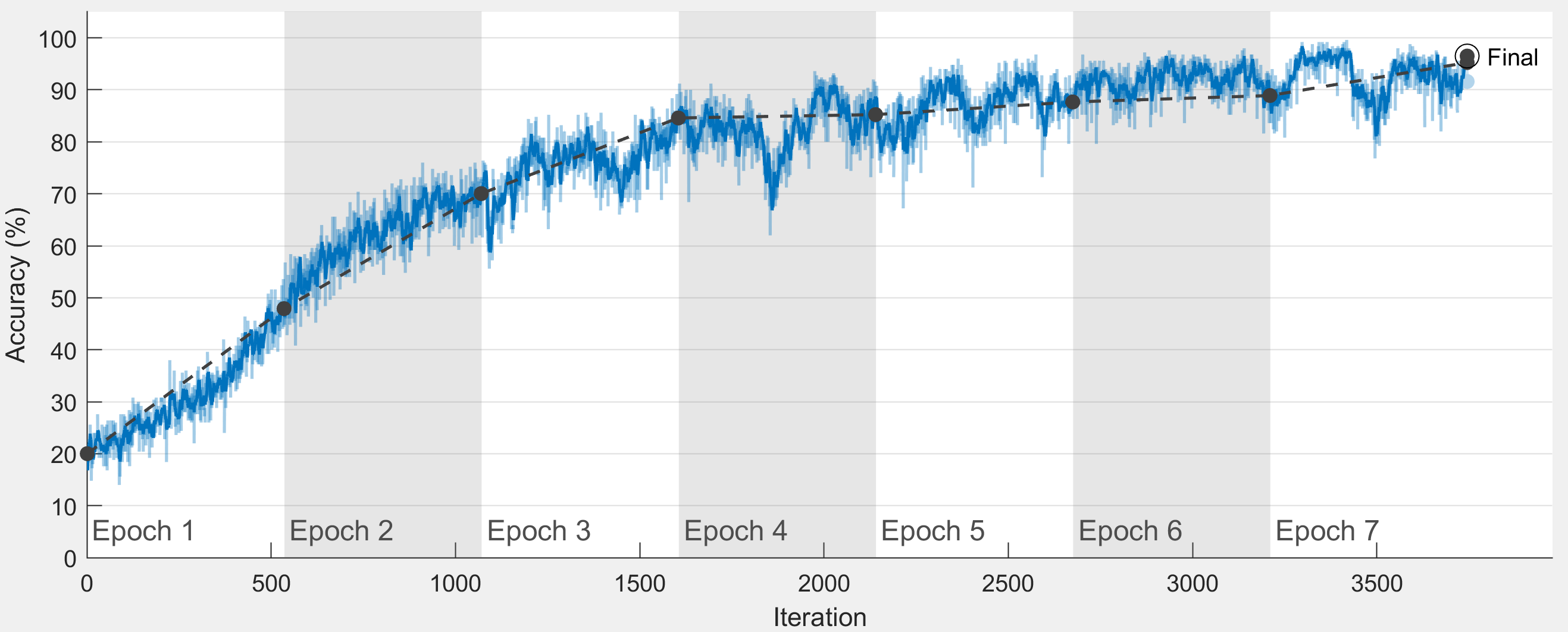}
    \label{subfig-1:train1}  } 
	\vspace{0.0001mm}
	\subfloat[Existing in-band only technique.]{
    \includegraphics[width=.85\columnwidth,height=.18\textheight]{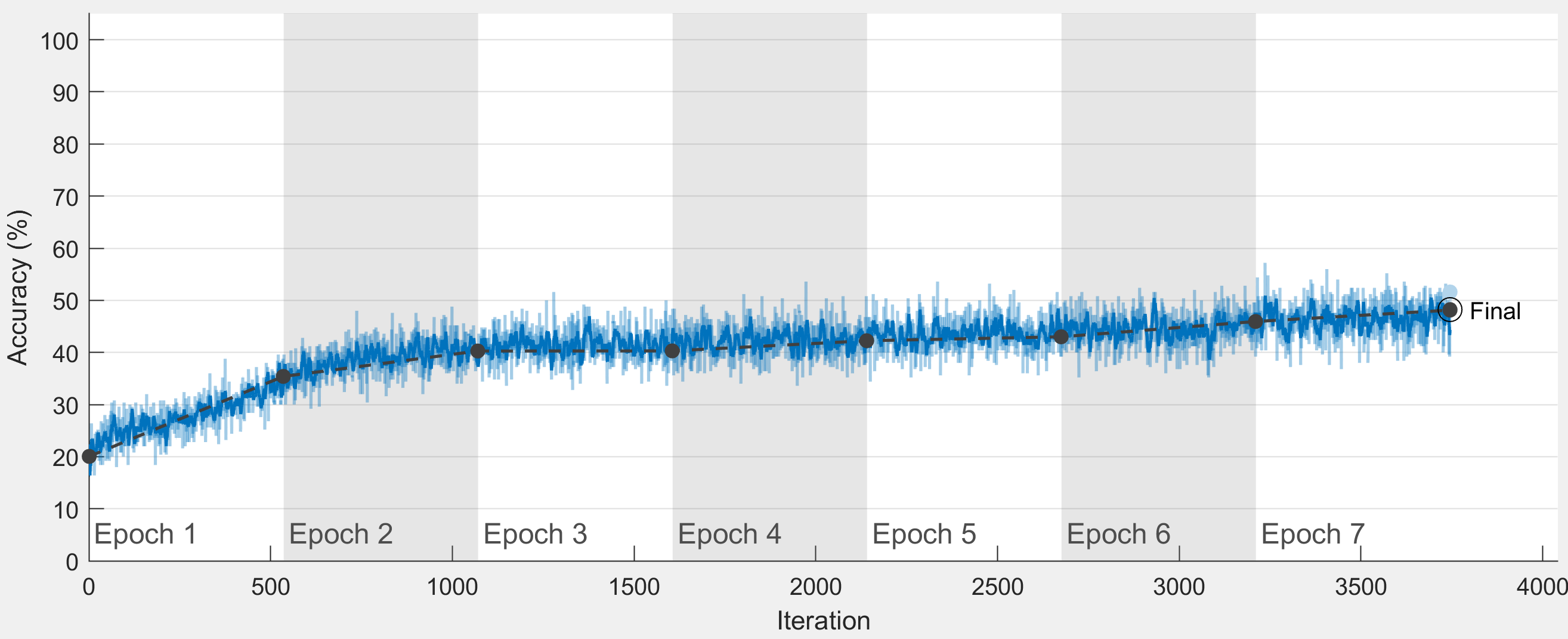}
    \label{subfig-2:train2}  } 
    \caption{Training and validation accuracy}
    \label{fig:trian}
\end{figure}

\comment{
\begin{figure}
\begin{center}
\begin{minipage}[b]{.6\columnwidth}
\centering
\includegraphics[width=\columnwidth]{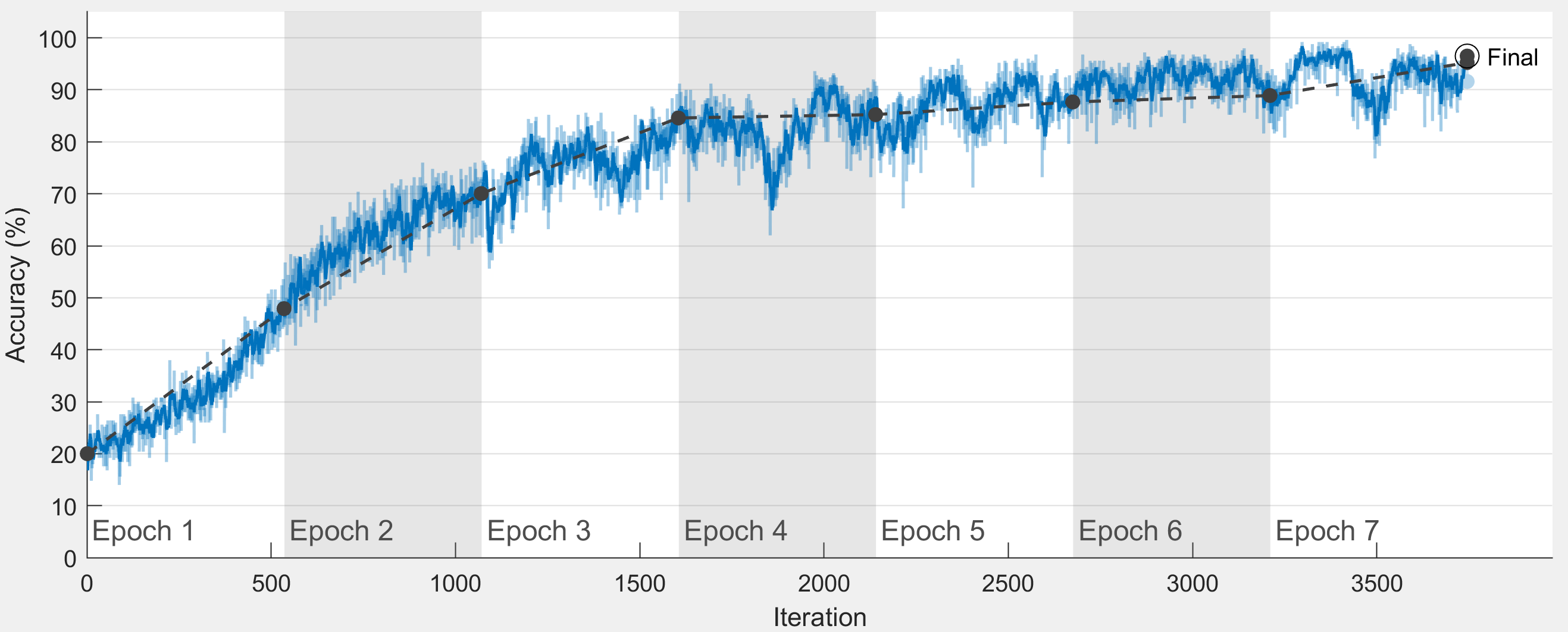}
\caption{Training Accuracy: In-band $\And$ Out-of-band information.}
\label{fig:figure12}
\end{minipage}
\begin{minipage}[b]{.6\columnwidth}
\centering
\includegraphics[width=\columnwidth]{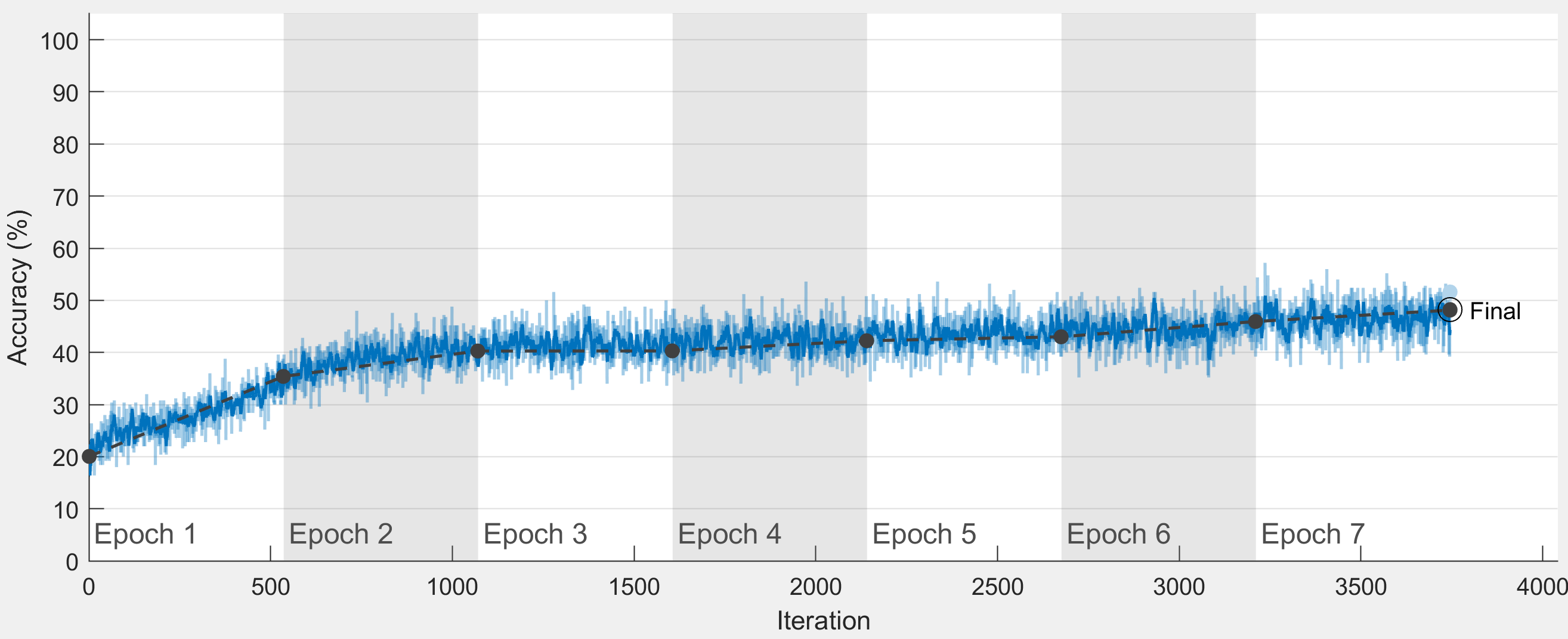}
\caption{Training Accuracy: In-band information only.}
\label{fig:figure13}
\end{minipage}
\end{center}
\end{figure}
}

We divided the generated frames into $80$\% training, $10$\% validation, and $10$\% testing. As we can see from Fig.~\ref{fig:trian}, the training accuracy (blue curve) of the proposed technique outperforms the traditional classifier that uses in-band information only. 
Our experiments show that the out-of-band additional processing exploited in our proposed technique does not incur an increase in the computation time of the method; the running times of the reported results are $97.38$ and $96.35$ minutes for the in-band only technique and the proposed in-band and out-of-band technique, respectively. Also, from the validation accuracy (the black dotted line in the figure), we can infer that our model does not suffer from overfitting.

The confusion matrices results depicted in Fig.~\ref{fig:conf_mat} show that the proposed technique achieves substantially higher classification accuracy than the in-band only technique. The testing accuracy obtained under the proposed technique across the five tested devices is \textbf{96.2\%} whereas that obtained under the in-band only approach is only \textbf{48.6\%}. 
It is worth mentioning that similar results are also obtained when considering the 8-PSK modulation scheme as opposed to the 16QAM scheme, though these results are presented in this paper.

Our technique achieves much higher accuracy because it leverages, in addition to the in-band distortion information already exploited by the prior methods, out-of-band distortion information caused by the different radio hardware components, which, as explained in the previous sections, provide unique device signatures that lead to substantial increase in device separability.


Another point that is also worth mentioning is that our experiments indicated that this accuracy gab between our proposed technique and the prior in-band only method is inversely proportional to the hardware impairments variability among devices, meaning that both techniques enjoy high classification accuracy when the devices exhibit relatively high hardware impairments. However, we strongly argue that as technology advancements continue to reduce such impairments, the variability among the hardware impairments across different devices will continue to shrink, making the reliance on only in-band information for device classification inefficient. Our proposed technique leveraging out-of-band distortion in addition to in-band information becomes in this case increasingly compelling and suitable for providing high device separability performance.


\comment{
\begin{figure}
	\centering
	\subfigure[Proposed in-band and out-of-band technique]{
    \includegraphics[width=1\columnwidth,height=.2\textheight]{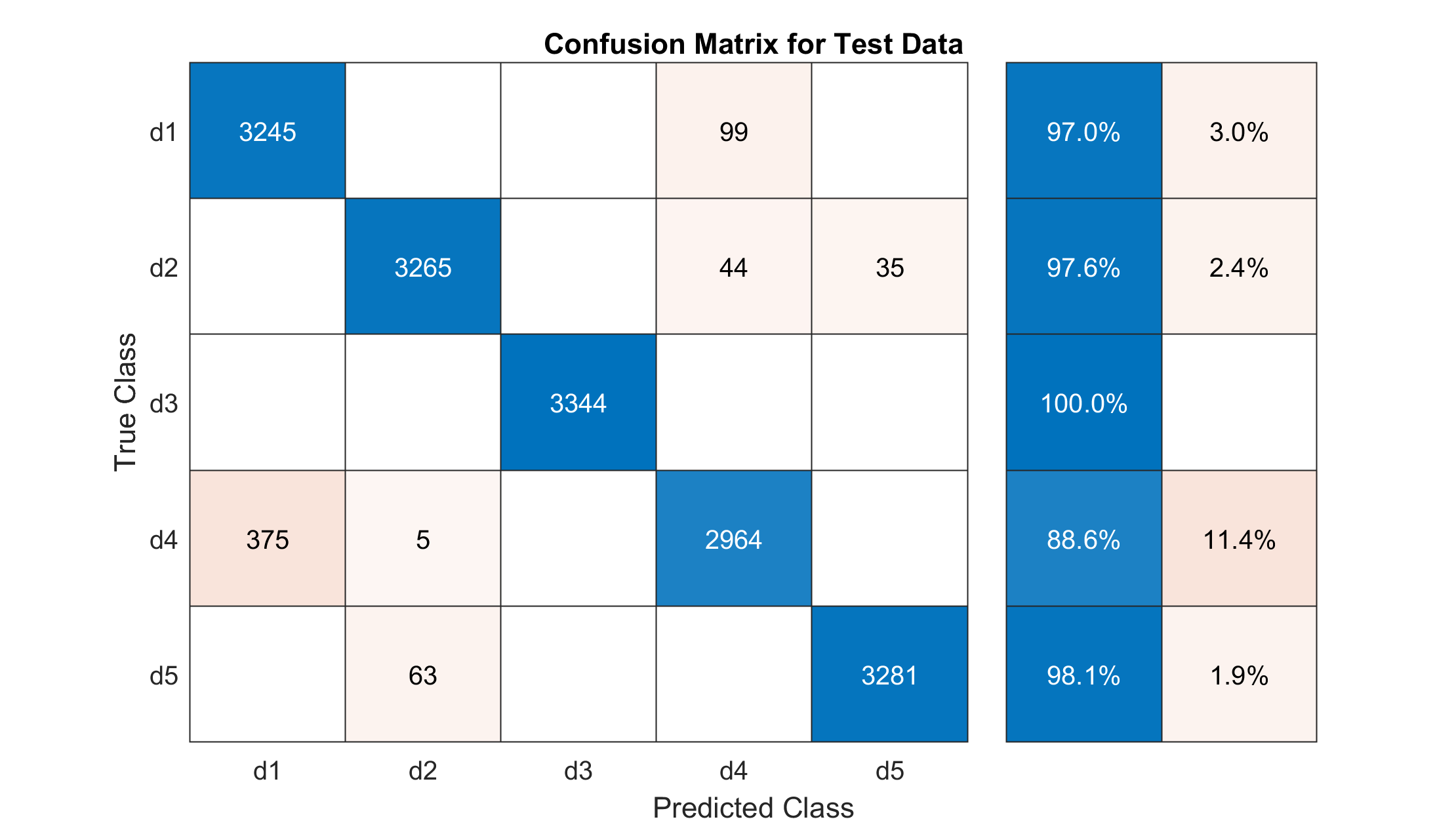}
    \label{subfig-1:conf1}  } 
	\subfigure[In-band only technique]{
    \includegraphics[width=1\columnwidth,height=.2\textheight]{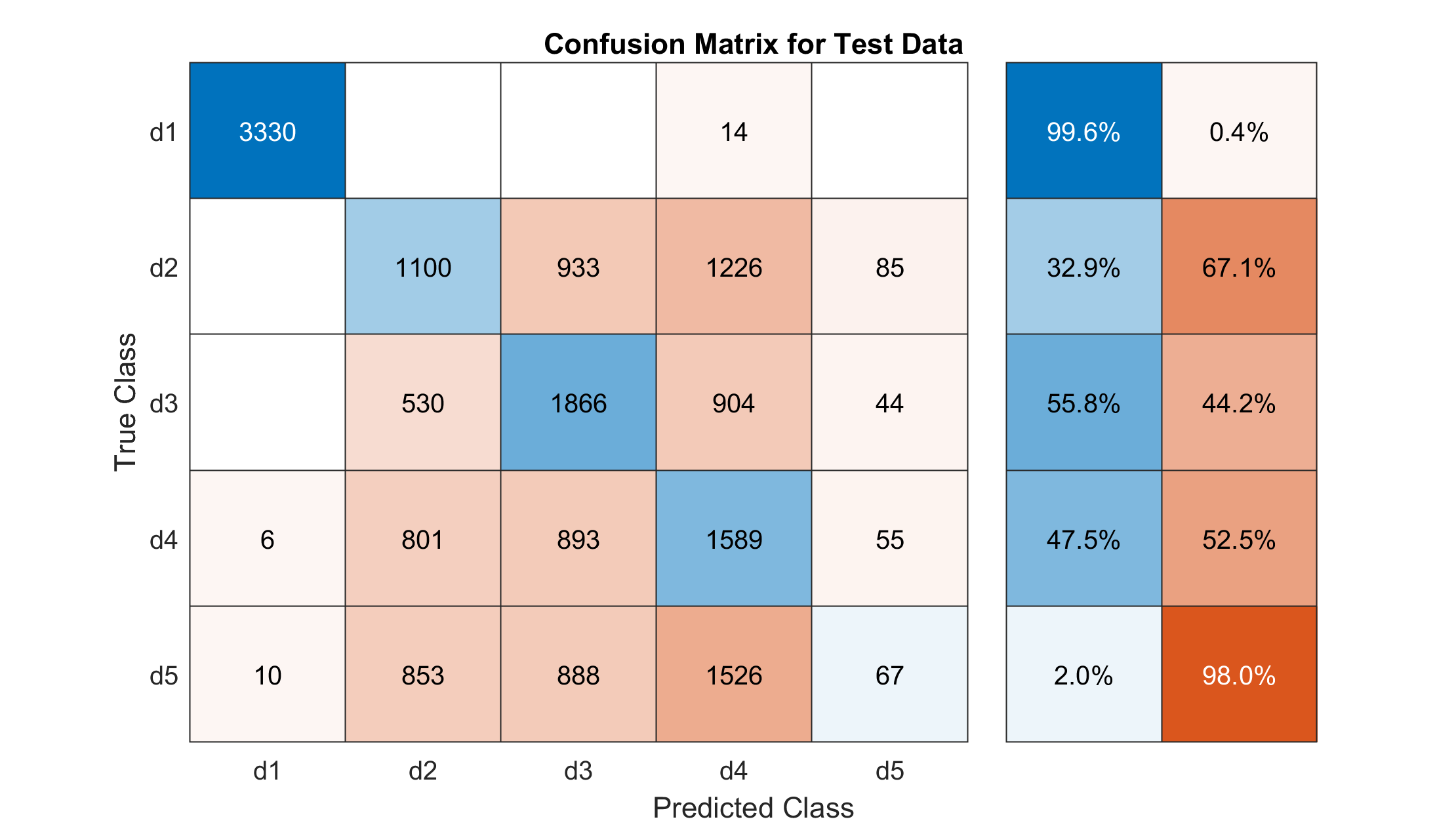}
    \label{subfig-2:conf2}  } 
    \caption{Confusion matrices}
    \label{fig:conf_mat}
\end{figure}
}

\begin{figure}
    \centering
     \subfloat[Our proposed in-band and out-of-band technique.\label{subfig-1:conf1}]{%
       \includegraphics[width=0.45\textwidth]{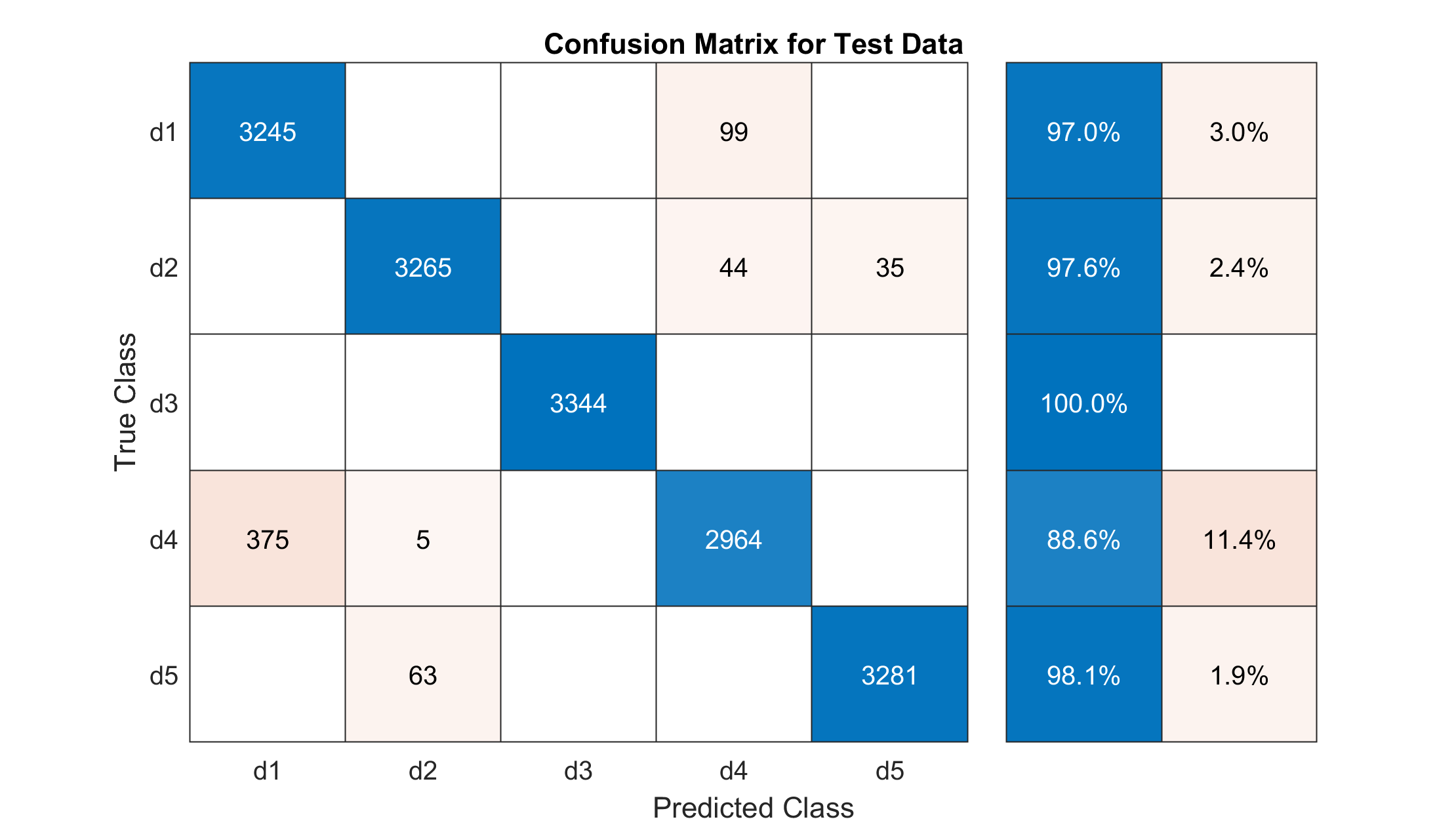}
     }
    \vspace{0.00001cm}
     \subfloat[In-band only technique.\label{subfig-2:conf2}]{%
       \includegraphics[width=0.45\textwidth]{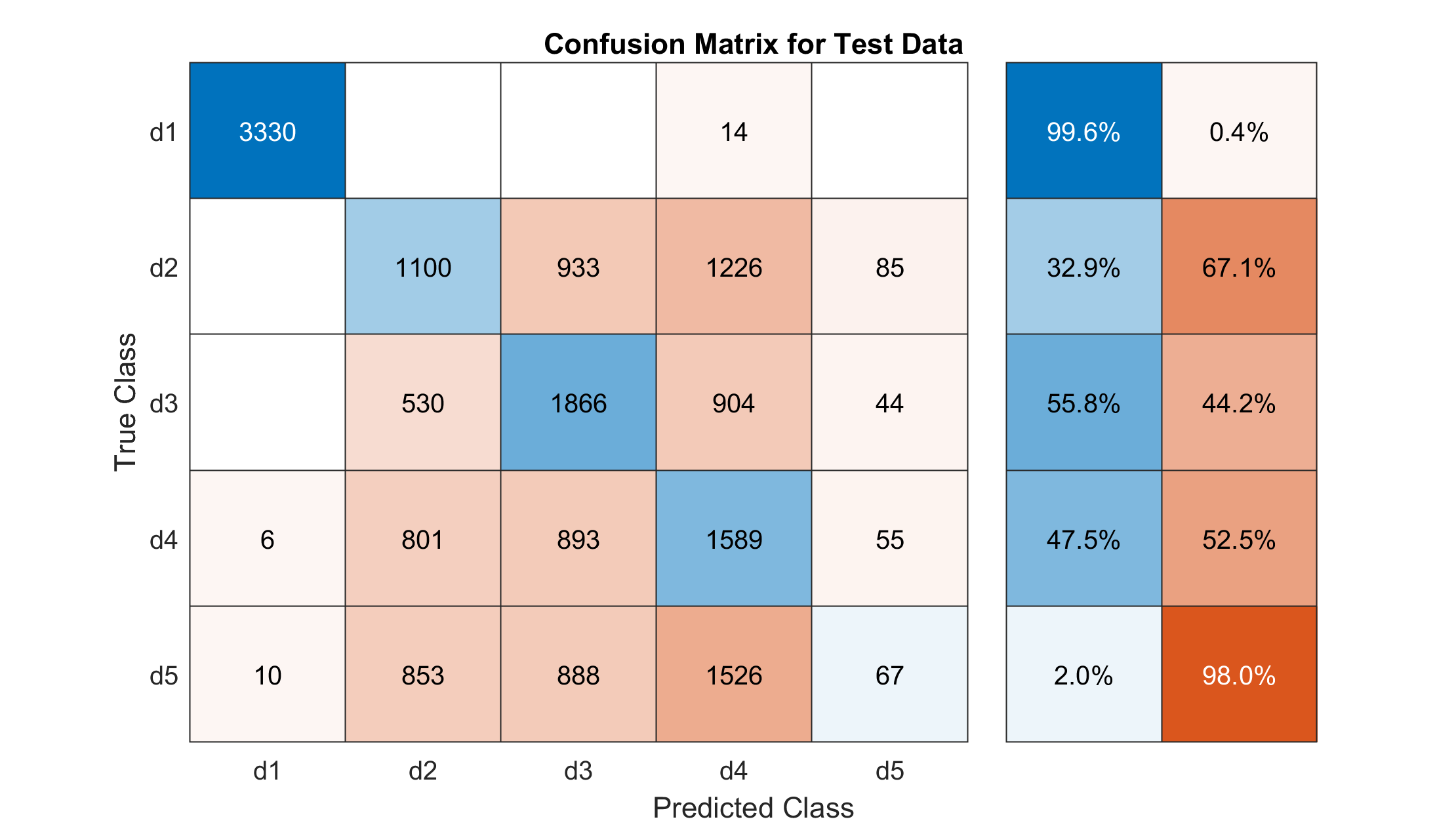}
     }
     \caption{Confusion matrices}
     \label{fig:conf_mat}
\end{figure}




\section{Conclusion}
\label{Conclusion}
In this paper, we proposed a scalable and non-clonable device classification technique that exploits both the in-band and out-of-band signal information via tuned CNN architecture classifier to provide robust and efficient device identification and classification. We presented the models and the impact of the main hardware RF transmitter impairments in considerable depth and insight, with more emphasis on the out-of-band signal distortions and their potentials and contributions to providing unique device signatures and features that can increase devices' separability and classification.
We evaluated the proposed technique on a simulated transmission chain of 5 devices impaired with almost identical hardware impairments. Experimental results showed that our proposed technique increases the device classification accuracy significantly, especially in realistic scenarios where the variability of hardware impairment values among the different devices is insignificant, which is the case of high-end, high-performance radios.

\bibliographystyle{IEEEtran}
\bibliography{conference}
\vspace{12pt}

\end{document}